\documentclass[namedreferences]{solarphysics}
\usepackage[hyperref,optionalrh]{spr-sola-addons} 
\usepackage{graphicx}        
\usepackage{amssymb}        
\usepackage{color}           
\usepackage{breakurl}        
\usepackage{tabularx}
\usepackage{multirow}
\usepackage{wasysym}

\usepackage{multirow}
\usepackage{amsmath}

\newcommand\at[2]{\left.#1\right|_{#2}}

\chardef\us=`\_

\begin{document}

\begin{article}

\begin{opening}

\title{Mixed properties of slow magnetoacoustic and entropy waves in a plasma with heating/cooling misbalance}

\author[addressref={aff1,aff2},corref,email={dimanzav@mail.ru}]{\inits{D.I.}\fnm{D.}~\lnm{Zavershinskii}\orcid{0000-0002-3746-7064}}

\author[addressref={aff3,aff4},corref,email={}]{\inits{D.Y.}\fnm{D.}~\lnm{Kolotkov}\orcid{0000-0002-0687-6172}}
\author[addressref={aff1,aff2},corref,email={}]{\inits{D.S.}\fnm{D.}~\lnm{Riashchikov}\orcid{0000-0001-7143-2968}}
\author[addressref={aff2},corref,email={}]{\inits{N.E.}\fnm{N.}~\lnm{Molevich}\orcid{0000-0001-5950-5394}}

%

\runningauthor{Zavershinskii et al.}
\runningtitle{Mixed properties of slow magnetoacoustic and entropy waves}

\address[id=aff1]{Department of Physics, Samara National Research University, Moscovskoe sh. 34, Samara, 443086, Russia}
\address[id=aff2]{Department of Theoretical Physics, Lebedev Physical Institute, Novo-Sadovaya st. 221, Samara, 443011, Russia}
\address[id=aff3]{Centre for Fusion, Space and Astrophysics, Department of Physics, University of Warwick, Coventry, CV4 7AL, United Kingdom}
\address[id=aff4]{Institute of Solar-Terrestrial Physics SB RAS, Irkutsk, 664033, Russian Federation}

\begin{abstract}
The processes of the coronal plasma heating and cooling were previously shown to significantly affect the dynamics of slow magnetoacoustic (MA) waves, causing amplification or attenuation, and also dispersion. However, the entropy mode is also excited in such a thermodynamically active plasma and is affected by the heating/cooling misbalance too. This mode is usually associated with the phenomenon of coronal rain and formation of prominences. Unlike the adiabatic plasmas, the properties and evolution of slow MA and entropy waves in continuously heated and cooling plasmas get mixed. Different regimes of the misbalance lead to a variety of scenarios for the initial perturbation to evolve. In order to describe properties and evolution of slow MA and entropy waves in various regimes of the misbalance, we obtained an exact analytical solution of the linear evolutionary equation. Using the characteristic timescales and the obtained exact solution, we identified regimes with qualitatively different behaviour of slow MA and entropy modes. For some of those regimes, the spatio-temporal evolution of the initial Gaussian pulse is shown. In particular, it is shown that slow MA modes may have a range of non-propagating harmonics. In this regime, perturbations caused by slow MA and entropy modes {in a low-$\beta$ plasma} would look identically in observations, as non-propagating disturbances of the plasma density (and temperature) either growing or decaying with time.
We also showed that the partition of the initial energy between slow MA and entropy modes depends on the properties of the heating and cooling processes involved. The obtained exact analytical solution could be further applied to the interpretation of observations and results of numerical modelling of slow MA waves in the corona and the formation and evolution of coronal rain.
\end{abstract}

%

\end{opening}

%
 \section{Introduction}
 	\label{s:Introduction} 




In a  series of recent works \citep{2021SoPh..296...20P, 2020arXiv201110437D, 2020A&A...644A..33K, 2019A&A...628A.133K,2019A&A...624A..96C, 2019PhPl...26h2113Z}, the thermodynamic activity of the solar corona, i.e. a wave-induced interplay between plasma heating and cooling processes, was evidently shown to be among the most important physical processes affecting the dynamics of magnetoacoustic waves and oscillations in the corona. More specifically, a compressive wave violating the coronal thermal equilibrium via perturbations of the local plasma parameters can experience a feedback from these unbalanced coronal heating and cooling processes, which is known as the phenomenon of thermal misbalance \citep{1965ApJ...142..531F,Molevich88,2017ApJ...849...62N}. In a broad range of typical physical conditions in the Sun's corona, the characteristic timescales of this thermal misbalance were shown to be about several minutes which coincide with the oscillation periods of magnetoacoustic waves ubiquitously present in the coronal plasma structures \citep[see e.g.][for the most recent comprehensive review]{2020ARA&A..58..441N}. In particular, the essentially compressive slow-mode magnetoacoustic waves are confidently observed as propagating or standing disturbances in various coronal plasma non-uniformities such as, for example, polar plumes, quiescent and active region loops, and have oscillation periods ranging from a few to a few tens of minutes and comparable damping times \citep{2009SSRv..149...65D, 2011SSRv..158..267B, 2011SSRv..158..397W, 2016GMS...216..419B, 2016GMS...216..395W, 2019ApJ...874L...1N, 2021SSRv..217...34W}. As such, slow waves are considered to be strongly affected by the thermal misbalance process.

The direct observations and theoretical modelling of coronal slow waves are extensively used for seismological probing the coronal plasma. In particular, the effects of parallel thermal conduction and compressive viscosity, leading to a frequency- and temperature-dependent damping of slow waves and phase shifts between density and temperature perturbations, were measured in e.g. \citet{2014ApJ...789..118K, 2015ApJ...811L..13W, 2018ApJ...860..107W, 2019FrASS...6...57S} and modelled in e.g. \citet{2002ApJ...580L..85O, 2003A&A...408..755D, 2005A&A...436..701S, 2009A&A...494..339O, 2016ApJ...826L..20R, 2016ApJ...820...13M}. An effective coronal polytropic index and its dependence on temperature were inferred seismologically with slow waves in \citet{2011ApJ...727L..32V} and \citet{2018ApJ...868..149K}, respectively.
Robust and reliable methods for measuring the apparent propagation speed of slow waves in legs of long fan-like loops in active regions, as an important seismological tool, were designed by \citet{2012A&A...543A...9Y}. Using a combination of the theory of the perturbed thermal equilibrium and observations of slow waves in long-lived coronal plasma structures, \citet{2020A&A...644A..33K} seismologically constrained the parameters of the unknown coronal heating function. Likewise, \citet{2019ApJ...884..131R} used observations of large-amplitude quasi-periodic pulsations associated with slow waves and a comprehensive theoretical modelling to diagnose duration of the impulsive coronal heating events.

The feedback from the wave-induced thermal misbalance on the slow wave dynamics is manifested, in particular, in the wave damping or amplification. Thus, \citet{2019A&A...628A.133K} demonstrated three possible regimes of the standing slow wave evolution with enhanced damping (with respect to that caused by other dissipative processes, e.g. parallel thermal conduction) or suppressed damping and even amplification (over-stability) of the wave through the effective gain of energy from the heating source. It was shown that the observed rapid damping of standing slow oscillations in hot and dense coronal loops could be readily reproduced with a reasonable choice of the heating function. This study was generalised and extended in \citet{2020arXiv201110437D} for the effects of non-zero plasma-$\beta$ and broader range of coronal conditions. In the over-stable regime, the effects of finite amplitude become important, leading to the distortion of the wave front and/or formation of autowave (self-sustained) shock pulses \citep{2020PhRvE.101d3204Z, 2017ApJ...849...62N, Molevich2016191, 2013TePhL..39..676Z, 2010PhPl...17c2107C, 2000ApJ...528..767N}. The nonlinear evolution of slow magnetoacoustic waves in the corona was also addressed by e.g. \citet{2002ApJ...580L..85O}, \citet{2006RSPTA.364..485R}, \citet{2008ApJ...685.1286V}, \citet{2015A&A...573A..32A}, and \citet{2019ApJ...886....2W}. 

\citet{2019PhPl...26h2113Z} demonstrated another effect of the heating/cooling misbalance that is effective dispersion of slow waves. This additional misbalance-caused dispersion is manifested through the dependence of the wave phase and group speeds and effective polytropic index on the wave frequency; is fully attributed to the existence of the characteristic timescales of the misbalance process; and is not connected with the geometrical dispersion of magnetoacoustic waves, traditionally considered in the corona. Moreover, the slow-mode wave increment/decrement (i.e. damping/growth time or length, respectively) caused by the misbalance acquires the frequency dependence too, which is functionally different to that caused by the parallel thermal conduction and viscosity. In particular, accounting for this frequency-dependent damping by the misbalance in the model allowed \citet{2021SoPh..296...20P} to better match the observed relationship between the oscillation period and damping time of the fundamental standing slow waves. Combining the wave dispersion and frequency-dependent damping/amplification both associated with the effect of thermal misbalance, \citet{2019PhPl...26h2113Z} demonstrated formation of a quasi-periodic propagating slow magnetoacoustic wave train from the initially broadband aperiodic perturbation.

A particular regime, in which the wave evolves much faster/slower than the misbalance process does, corresponds to the weak/strong limits of the latter, respectively. Thus, in the regime of weak misbalance and low-$\beta$ plasma, the slow wave speed reduces to the standard adiabatic sound speed determined by the standard adiabatic index and the plasma temperature. In the regime of dominating misbalance, the wave becomes strongly non-adiabatic with a modified propagation speed and a new value of the effective polytropic index determined  by the properties of the heat-loss processes. In both these regimes, the wave speeds are frequency-independent. This is similar to the theory of the lower and higher limits of the parallel thermal conduction \citep[see e.g. Sec.~4.1 in][]{2014ApJ...789..118K}, for which the slow wave speed varies between the adiabatic and isothermal values of the sound speed, respectively. A more general model of slow waves, including the combined effects of the weak/strong misbalance and lower/higher limits of thermal conduction that for uniformity was referred to as regimes of weak/strong non-adiabaticity, was considered by \citet{2020arXiv201110437D}.

Another natural magnetohydrodynamic (MHD) process efficiently perturbing the coronal thermal equilibrium and strongly affected by the back-reaction of this perturbation is the entropy wave. It is a non-propagating compressive mode \citep[see e.g.][]{2011A&A...533A..18M,2007AstL...33..309S}, either growing or decaying in response to the violation of the balance between plasma heating and cooling processes \citep{1965ApJ...142..531F}. In particular, the radiative instability of the entropy mode could lead to rapid condensations of the coronal plasma and formation of coronal rain \citep[see e.g.][and references therein, for the most recent review]{2020PPCF...62a4016A} or prominences \citep[e.g.][]{2017ApJ...845...12K}. Hence, in a continuously heated and cooling plasma of the solar corona, the properties and evolution of entropy waves and slow magnetoacoustic waves could get mixed through the mechanism of thermal misbalance, even in the linear regime.
The question of mixed properties of MHD waves in the solar corona is usually considered in the context of fast magnetoacoustic kink and torsional Alfv\'en waves in coronal loops \citep[see e.g.][for a recent work]{2019FrASS...6...20G}, while the mixed properties of slow magnetoacoustic and entropy waves in the coronal plasma with heating/cooling misbalance has not been addressed in the previous theoretical or observational works yet.

In this paper, we perform a comprehensive analytical treatment of slow magnetoacoustic and entropy waves, simultaneously excited and evolving in the linear regime in a thermodynamically active plasma of the solar corona, by obtaining an exact analytical solution of the evolutionary equation (see Sec.~\ref{sec:wave_eq}--\ref{Subsec 4_1}). {{This exact analytical solution for the linear wave dynamics in a plasma with thermal misbalance is derived for the first time in this work.}}
Unlike the previous works \citep[e.g.][]{2019A&A...628A.133K, 2019PhPl...26h2113Z} that considered the characteristic timescales of the thermal misbalance to be strictly positive to avoid the instability of the entropy mode, {{the novel element of the present work is that we do not use this assumption}}. Indeed, the estimations of the characteristic misbalance time for typical coronal conditions \citep[see e.g. Table~2 and Fig.~3 in][]{2020A&A...644A..33K} have shown that it could be both positive and negative, depending on the properties of the heat-loss function. Taking these negative times into account in this work allows us, in particular, to reveal the range of non-propagating slow magnetoacoustic harmonics, which are thereby not obviously distinguishable from the harmonics of the entropy wave. {{Revealing a specific regime of thermal misbalance in which slow magnetoacoustic harmonics may become non-propagating and thus possess properties similar to those of the entropy waves is another new result of this work.}} Using the obtained analytical solution, we describe and demonstrate different scenarios of the slow and entropy wave spatio-temporal evolution (Sec.~\ref{Subsec 4_2}). We stress that the individual dynamics of those waves cannot be analysed by solving the original evolutionary equation numerically which allows for obtaining the dynamics of their superposition only.
In Sec.~\ref{Subsec 4_3}, we demonstrate partition of the energy of the initial perturbation between slow and entropy waves, as another manifestation of their mixed properties. For example, this issue cannot be resolved from the analysis of the dispersion relation, and requires an exact solution of the evolutionary equation obtained in this work. {{The dependence of the partition of the initial perturbation energy between slow magnetoacoustic and entropy modes on the coronal heating and cooling processes is demonstrated for the first time in this work.}}
The discussion of the results and conclusions are given in Sec.~\ref{sec:summary}.

\section{Governing wave equation}
\label{sec:wave_eq}
The linear dynamics of slow magnetoacoustic (MA) and entropy waves in a uniform along the field plasma with the heating/cooling misbalance is described by the following evolutionary equation, derived in the infinite field approximation by \citet{2019PhPl...26h2113Z},
\begin{equation}\label{step_3}
	\frac{\partial^3 \rho_1}{\partial t^3} -  c_{S}^2 \frac{\partial^3 \rho_1}{\partial t \partial z^2}
	=\frac{\kappa}{\rho_0 C_\mathrm{V}}\left(\frac{\partial^4 \rho_1}{\partial z^2\partial t^2} - c_{S0}^2 \frac{\partial^4 \rho_1}{\partial z^4} \right) -\frac{1}{\tau_2 }\left(\frac{\partial^2 \rho_1}{\partial t^2} - c_{SQ}^2  \frac{\partial^2 \rho_1}{\partial z^2}\right).
\end{equation}
The two terms on the RHS of Eq.~(\ref{step_3}) describe the effects of the field-aligned thermal conduction with the coefficient $\kappa$ and thermal misbalance on small-amplitude plasma density perturbations $\rho_1$. More specifically, $c_{S} =\sqrt{\gamma k_\mathrm{B}  T_0 /m} $ and $ c_{S0} =\sqrt{k_\mathrm{B}  T_0 /m} $ are the standard adiabatic (with the adiabatic index $\gamma$) and  isothermal sound speeds, respectively;  $c_{SQ} =\sqrt{ \gamma_\mathrm{Q} {k_\mathrm{B}  T_0}/{m}  }$ is the  sound speed of wave propagation in the regime of strong misbalance (when the second term on the RHS of Eq.~(\ref{step_3}) dominates), prescribed by the  effective polytropic index $\gamma_\mathrm{Q}\equiv {Q_{[P]T}}/{Q_{[\rho]T}} = {\gamma\tau_2}/{\tau_1}$ \citep{1974A&A....37...65H, Molevich88} and the characteristic timescales of the misbalance,
\begin{align}
\label{eq:tau1}
&\tau_1=C_\mathrm{P}/Q_{[P]T},\\
\label{eq:tau2}
&\tau_2=C_\mathrm{V}/Q_{[\rho]T}.
\end{align}
Here, $Q_{[P]T}$ and $Q_{[\rho]T}$ are the derivatives of the combined heat ($H$) and loss ($L$) function 
{	\begin{equation}\label{QFunc}
	Q(\rho, T)=L (\rho , T ) - H(\rho , T ),
	\end{equation} }
with respect to the plasma temperature $T$, taken at the constant gas pressure $P$ and density $\rho$, i.e.  $Q_{[\rho]T} = \left( \partial Q  / \partial T \right)_{\rho}$, $Q_{[P]T}  =  \left( \partial Q  / \partial T \right)_{P} = \left( \partial Q  / \partial T \right)_{\rho}   - (\rho_0/T_0)  \left( \partial Q  / \partial \rho \right)_{T}$. In this notation, $k_\mathrm{B}$ is the Boltzmann constant,  $m$ is the mean particle mass, and $C_\mathrm{P}$ and $C_\mathrm{V}$ are specific heat capacities. 

{{As an initial equilibrium, we consider a long-lived coronal plasma with density $\rho_0$ and temperature $T_0$, and with the radiative cooling and heating rates balancing each other, so that $Q_0(\rho_0, T_0)=0$. As seen from Eq.~(\ref{step_3}), the effects caused by the linear wave perturbation of such a thermal equilibrium (wave-induced thermal misbalance) are determined not by the sign and absolute value of $Q$, but by the sign and absolute values of its derivatives $Q_{[P]T}$ and $Q_{[\rho]T}$ or, equivalently, by the characteristic times $\tau_{1,2}$. Previous estimations by \citet{2020A&A...644A..33K} showed that for typical coronal conditions these timescales can be either positive or negative depending on a specific form of the heat-loss function $Q(\rho,T)$.}}
From the physical standpoint, those thermal misbalance timescales $\tau_1$ and $\tau_2$ are the main parameters determining the effect of the perturbed thermal equilibrium on the wave dynamics described by Eq.~(\ref{step_3}), which demonstrate how quickly the plasma restores its initial thermal equilibrium or becomes thermodynamically unstable.

The infinite magnetic field approximation used for obtaining Eq.~(\ref{step_3}) implies the magnetic field strength is high enough to neglect its perturbations by the slow MA wave and hence to consider the wave dynamics as one-dimensional, strictly along the field. Its applicability to slow MA waves in the solar corona was recently justified by \citet{2020arXiv201110437D}, for the magnetic field strength greater than 10~G in the quiescent loops and polar plumes and greater than 100~G in hot and dense loops in active regions. In these cases, the potential dependence of the unknown coronal heating function on the magnetic field strength was shown to have no effect on the wave dynamics. Hence, without loss of generality in our work this dependence is omitted in Eq.~(\ref{step_3}). Also, the effects of the gravitational stratification of the coronal plasma are missing in Eq.~(\ref{step_3}), that is consistent with the physical conditions in e.g. hot and dense loops for which the characteristic stratification scale height is known to be much greater than a typical loop height \citep[see e.g.][]{2018ApJ...860..107W, 2019ApJ...886....2W}. In this work, we use Eq.~(\ref{step_3}) originally obtained by \citet{2019PhPl...26h2113Z} as a starting point, without re-deriving it. Being the third-order equation {with respect to time}, Eq.~(\ref{step_3}) describes three wave modes, which are two slow MA modes and one entropy mode. 

In addition to the assumptions described above, the most recent seismological studies with slow waves revealed evidence of strong suppression of the parallel thermal conduction at least in some solar active regions \citep[see a series of works by][]{2015ApJ...811L..13W, 2018ApJ...860..107W, 2019ApJ...886....2W}. Likewise, \citet{2017A&A...600A..37N} obtained seismologically the effective adiabatic index of about 5/3 in the hot loop hosting a slow-mode oscillation, which could also be considered as an indirect evidence of a diminished efficiency of the parallel thermal conduction. Moreover, observations by \citet{2018ApJ...868..149K} showed the increase in the effective coronal polytropic index with temperature, that is also inconsistent with the theoretical prediction arising from the effect of thermal conduction.
Addressing these recent observational precedents of anomalously low thermal conduction, in this work we neglect the first term on the RHS of Eq.~(\ref{step_3}).

{{Thus, the main equation governing the linear evolution of slow MA and entropy waves in a plasma with heating/cooling misbalance and under the assumptions described above is}}
\begin{equation}\label{lin_evol_eq}
	\frac{\partial^3   \widetilde{\rho}}{\partial \tilde{t}^3} -  \gamma \frac{\partial^3 \widetilde{\rho}}{\partial \tilde{t} \partial \tilde{z}^2} = - \widetilde{\nu}_2 \left( \frac{\partial^2 \widetilde{\rho}}{\partial \tilde{t}^2} -  \gamma_Q \frac{\partial^2 \widetilde{\rho}}{ \partial \tilde{z}^2} \right).
\end{equation}
Here, we have introduced the dimensionless density perturbation $ \tilde{\rho}= \rho_{1}/  \rho_0$, coordinate $ \tilde{z}= z/L$, and time  $ \tilde{t}= t c_{S0} /{L}$, where $L$   is the characteristic spatial scale of the medium (for example, the loop length). Also, hereafter we use dimensionless characteristic frequencies $\tilde{\nu}_1$, $\tilde{\nu}_2$, and $\tilde{\nu}_{12}$, defined through the characteristic thermal misbalance timescales $\tau_1$ and $\tau_2$ as
\begin{equation}\label{chr_frq_2}
		\widetilde{\nu}_{1,2} = \frac{1}{ \widetilde{\tau}_{1,2}}  = \frac{L}{\tau_{1,2} c_{S0}},~~\widetilde{\nu}_{12} = \frac{\widetilde{\tau}_1-\widetilde{\tau}_2}{2 \widetilde{\tau}_2 \widetilde{\tau}_1  }.
\end{equation}
 
In Sec.~\ref{sec:parametric-analysis}, we demonstrate that depending on the ratio, sign and absolute values of the characteristic thermal misbalance times $\tau_{1,2}$ (or their dimensionless counterparts $\widetilde{\nu}_{1,2}$ and $\widetilde{\nu}_{12}$), the harmonics of slow MA and entropy waves may evolve differently. The further analysis is conducted for dimensionless quantities, hence the tilde sign is omitted.

\section{Parametric analysis}
\label{sec:parametric-analysis}

We obtain the solution for evolutionary equation (\ref{lin_evol_eq}) by separation of variables that is also known as the Fourier method. A perturbation of the plasma equilibrium state can, in general, be represented as a sum of the wave modes constituting it (in our case a sum of the entropy and slow MA modes). The Fourier method allows us to distinguish between the impact of these different physical modes, determine partition of the energy of the initial perturbation between them, and analyse their evolution separately. Moreover, this method allows us to analyse behaviour of the individual Fourier harmonics of those physical modes, which may grow or decay, and propagate or not propagate. The latter non-propagating behaviour retains across the whole spectrum of the perturbation for entropy waves and may occur in a specific interval of harmonics for slow MA waves. For example, in the discussed coronal plasma with thermal misbalance, some harmonics of the slow MA mode may grow and propagate, while other harmonics of the same mode may also grow but not propagate, if certain physical conditions are fulfilled. In this section, we derive those conditions explicitly for both the slow MA and entropy modes and link them with the characteristic timescales of the misbalance, $\tau_{1,2}$ (\ref{eq:tau1}), (\ref{eq:tau2}).

\subsection{Behaviour of the individual Fourier harmonics} \label{Subsec 3_1}
Applying the Fourier method, we search for the solution for Eq.~(\ref{lin_evol_eq}) in the form $\rho\left(z,t \right) = \varphi\left(z\right) \psi\left(t\right)  $. This substitution allows us to split  Eq.~(\ref{lin_evol_eq}) into two equations describing  dependence of the full solution on coordinate, $\varphi\left(z\right)$,  and on time, $\psi\left(t\right) $, respectively.

Equation describing the dependence of the perturbation on coordinate, $\varphi\left(z\right)$  has a form of the harmonic oscillator,
\begin{equation}\label{spatial_eq}
	\frac{d^2 \varphi}{ d z^2} + k^2  \varphi = 0,
\end{equation}
where $k^2$ {are the eigenvalues} that appear after the separation of variables in Eq.~(\ref{lin_evol_eq})  as $F_1(\psi, \psi_t', \psi_{tt}'', \psi_{ttt}''')=F_2(\varphi,\varphi_{zz}'')=-k^2$.
We assume that there are no mass flows at the boundaries, which implies that Neumann boundary conditions $ \partial \rho \left(0, t \right)  / \partial  z = \partial \rho \left(l, t \right)  / \partial z =0 $  or, equivalently, $d \varphi \left(0 \right)  /d  z = d \varphi \left(l \right)  / d z =0 $, are applied. Here, $l$ is the length of the medium normalised to the characteristic length scale  $L$. 

{The oscillatory solutions of Eq.~(\ref{spatial_eq}) exist only for eigenvalues $k^2>0 $. } In this case, the eigenvalues $k$ can be defined by the harmonic number $ n $  as
\begin{equation}\label{eigenvalue_eq}
	k = \frac{\pi n}{l}, n = 1,2,3,... 
\end{equation}
The eigenvalues (\ref{eigenvalue_eq}) correspond to the set of possible wavenumbers of the entropy and slow MA modes, which in turn defines the set of possible wavelengths $\lambda=2\pi/k$.
{{The quantisation of the wavenumbers $k$ in terms of the characteristic length of the medium $l$ implies the existence of a closed resonator (for example, the coronal loop), which would eventually allow the initially localised MA perturbation to form standing waves if it does not dissipate or leak out before getting reflected from the resonator boundary. On the other hand, the developed theory can be readily applied for description of propagating waves in open plasma structures too, if the wave travel time to the resonator boundary is longer than its lifetime or by the use of open boundary conditions. In this case, the set of wavenumbers $k$ would be continuous and fully prescribed by the driver.}}

 Solution of Eq.~(\ref{spatial_eq}) for $k^2>0$ and the chosen boundary conditions is well-known and gives us the spatial dependencies $\varphi\left(z\right)=\varphi_n\left(z\right)$ for $n>0$  of the full solution  $\rho\left(z,t \right)$. { The particular case with  $k^2=0$ ($n=0$)  corresponds to a non-oscillatory background of the full solution $\rho_{0}(z,t)$ with spatial dependence $\varphi_0\left(z\right)$. }

In order to describe temporal evolution of the entropy and slow MA harmonics with { wavenumbers $k^2 > 0$}, we consider equation for  $\psi\left(t\right) $, arising from the same separation of variables procedure. This equation is the third-order linear ordinary differential equation with constant coefficients written as
\begin{equation}\label{temporal_eq}
	\frac{d^3 \psi}{ d t^3} +\nu_2  \frac{d^2 \psi}{ d t^2} +  k^2(n) \gamma   \frac{d \psi}{ d t} + k^2(n) \gamma_Q \nu_2  \psi = 0.
\end{equation}
The solution of Eq. (\ref{temporal_eq}) may have different forms \citep[see e.g.][]{1995heso.book.....P}, depending on the type of roots of the following cubic algebraic equation,
\begin{equation}\label{cubic_eq}
	\omega^3 - i \nu_2  \omega^2 -  k^2(n) \gamma   \omega + i k^2(n) \gamma_Q \nu_2 = 0,
\end{equation}
obtained by writing $d/d\,t \to i\omega$ in Eq. (\ref{temporal_eq}). The cubic equation with complex coefficients (\ref{cubic_eq}) coincides with the general dispersion relation for slow and entropy modes  in the plasma with thermal misbalance, derived in e.g. \citet{2019PhPl...26h2113Z}, \citet{2019A&A...628A.133K}, \citet{Ryashchikov2017416} and \citet{1965ApJ...142..531F}.  Thus,  roots of Eq. (\ref{cubic_eq}) can be considered as complex frequencies $\omega_{1,2,3}$  of entropy and slow MA harmonics, corresponding to real wavenumber $k$.
{{The asymmetry in the temporal and spatial dependencies (powers of $\omega$  and $k$ ) described by Eq.~(\ref{cubic_eq}) can be explained by the fact that out of its all three possible solutions only two slow MA modes can propagate. The phase speed of the entropy mode is always equal to zero.}}

The discriminant $\Delta$ of Eq. (\ref{cubic_eq}) is 
\begin{equation}\label{discriminant_eq}
\Delta=-108(R^3+U^2), 
\end{equation}
where  $R$ and $U$ are real coefficients,
\begin{equation}\label{notation_discr_eq}
R = \frac{3 k^2 \gamma - \nu_2^2 }{9},~~U = \frac{ 2 \nu_2^3 - 9 \nu_2 k^2 \gamma +27 k^2 \gamma_Q \nu_2 }{ 54}.\nonumber
\end{equation}
Thus, as the discriminant  $\Delta$ (\ref{discriminant_eq}) is purely real-valued, frequencies $\omega_{1,2,3}$ may have the following types for given $k$:
\begin{itemize}
	\item \textit{Case 1} with {one purely imaginary} root $\omega_1 $ and {two complex conjugate} roots  $\omega_{2,3}$. This case is physically equivalent to the existence of one non-propagating (entropy) harmonic and two propagating (slow MA) harmonics.
	\item \textit{Case 2} with all  {three} roots  $\omega_{1,2,3}$ being purely imaginary. In this case,  not only the entropy harmonic, but also two slow MA harmonics become non-propagating.
\end{itemize}
Occurrence of \textit{Cases 1,2} directly depends on the sign and absolute values of the constant coefficients in Eq. (\ref{cubic_eq}). In our problem, they are defined not only by the harmonic number $n$, but, more importantly, by the properties of the heat-loss function $ Q (\rho , T ) $  through the quantities $ \gamma_Q = {\gamma\tau_2}/{\tau_1}$ (\ref{eq:tau1}),~(\ref{eq:tau2}) and $\nu_2 = {L}/(\tau_{2} c_{S0}) $ (\ref{chr_frq_2}). In other words, for some chosen heat-loss function $ Q (\rho , T ) $ providing fixed values of the timescales $\tau_1$ and $\tau_2$ (and their combinations $\gamma_Q$ and $\nu_2$),  \textit{Case 1} (with  one non-propagating and two propagating harmonics) does not necessarily hold true for all harmonic numbers $n$ in the perturbation spectrum. Thus, {the possibility for certain MA harmonics excited in a thermodynamically active plasma of the solar corona to propagate or not propagate directly depends on the coronal heat-loss function $ Q (\rho , T ) $}.


According to the Cardano's formula,  roots of dispersion relation (\ref{cubic_eq}) can be written as
\begin{align}\label{roots_eq} 
	&\omega_1  = i\left(-\frac{\nu_2}{3}+A+B\right),\nonumber \\ 
	&\omega_2 =   \frac{A-B}{2} \sqrt{3}  -i\left(\frac{\nu_2}{3}+\frac{A+B}{2}\right),\\
	&\omega_3 =  - \frac{A-B}{2} \sqrt{3} -i\left(\frac{\nu_2}{3}+\frac{A+B}{2}\right),  \nonumber
\end{align}
where 
\begin{equation}\label{notation_roots_eq}
	A = \sqrt[3]{-U + \sqrt{-\Delta/108}} ,~~B = - R / A.\nonumber
\end{equation}


We can discriminate between the aforesaid \textit{Cases 1,2}, using discriminant $\Delta$ (\ref{discriminant_eq}) and roots $\omega_{1,2,3}$ (\ref{roots_eq}) for given harmonic number $n$. { Thus, \textit{Case 1} occurs  for $\Delta<0 $. Taking a real-valued cubic root in the expression for the coefficient $A$ in Eqs. (\ref{roots_eq}), we can  determine the type of roots $\omega_{1,2,3}$ and uniquely associate them with the physical wave modes. In this case, the frequency $\omega_1$ is purely imaginary and  the conjugate frequencies $\omega_{2,3}$ are complex.} This corresponds to a non-propagating entropy harmonic  with the number $n$, which has zero real part, $\omega_{ER}=0$ and non-zero imaginary part, $ \omega_{EI} \ne 0$ (i.e. increment/decrement) of the frequency $\omega_1$,
\begin{equation}\label{im_frq_ent_eq}
	\omega_{EI} = -\frac{\nu_2}{3}+A+B. 
\end{equation}
In addition, there are two slow MA harmonics propagating in the opposite directions with the same number $n$. They have non-zero real part,  $\omega_{AR} \ne 0$   and  non-zero  imaginary part, $\omega_{AI} \ne 0$ (i.e. increment/decrement) of the complex conjugate frequencies $\omega_{2,3}=\pm\omega_{AR}-i\omega_{AI}$.  From  Eq.~(\ref{roots_eq}), those MA real and imaginary parts are
\begin{equation}\label{re_im_frq_ac_eq}
	\omega_{AI} =\frac{\nu_2}{3} + \frac{A+B}{2},~~\omega_{AR}=\frac{A-B}{2} \sqrt{3}.
\end{equation}

\textit{Case 2} with {three purely imaginary} roots $\omega_{1,2,3}$ (\ref{roots_eq}), corresponding to one non-propagating entropy harmonic and two non-propagating slow MA harmonics, occurs for $\Delta>0 $. {In this case, all values of the cubic root in the expression for the coefficient $A$ are complex. It can be shown that for this case  both slow MA and entropy harmonics have zero real part of the frequency  $ \omega_{AR} = \omega_{ER} = 0 $.} However, all modes have non-zero imaginary parts $\omega_{AI}, \omega_{EI} \ne 0 $ (i.e. increment/decrements), coinciding with the imaginary roots $\omega_{1,2,3}$ (\ref{roots_eq}). In this regime, perturbations of the plasma density and/or temperature caused by slow MA and entropy harmonics with the same wavenumber $k(n)$ would look identically, as a non-propagating disturbance either growing or decaying with time. In other words, it is not always obvious to distinguish between the physical modes of a slow MA or entropy nature in the set of roots $\omega_{1,2,3}$ (\ref{roots_eq}) in \textit{Case 2}. On the other hand, for varying properties of the heat-loss function  $ Q (\rho , T ) $, i.e. different values of the parameters $\tau_1$ and $\tau_2$, one of those imaginary roots can change sign (causing the corresponding physical mode to be stable or unstable) independently of the two other imaginary roots. 
In the following analysis, we use this property as a rule of thumb to differentiate between slow MA and entropy modes among the roots $\omega_{1,2,3}$ (\ref{roots_eq}) in \textit{Case 2}.

\begin{figure}
	\begin{center}
		\includegraphics[width=0.49\linewidth]{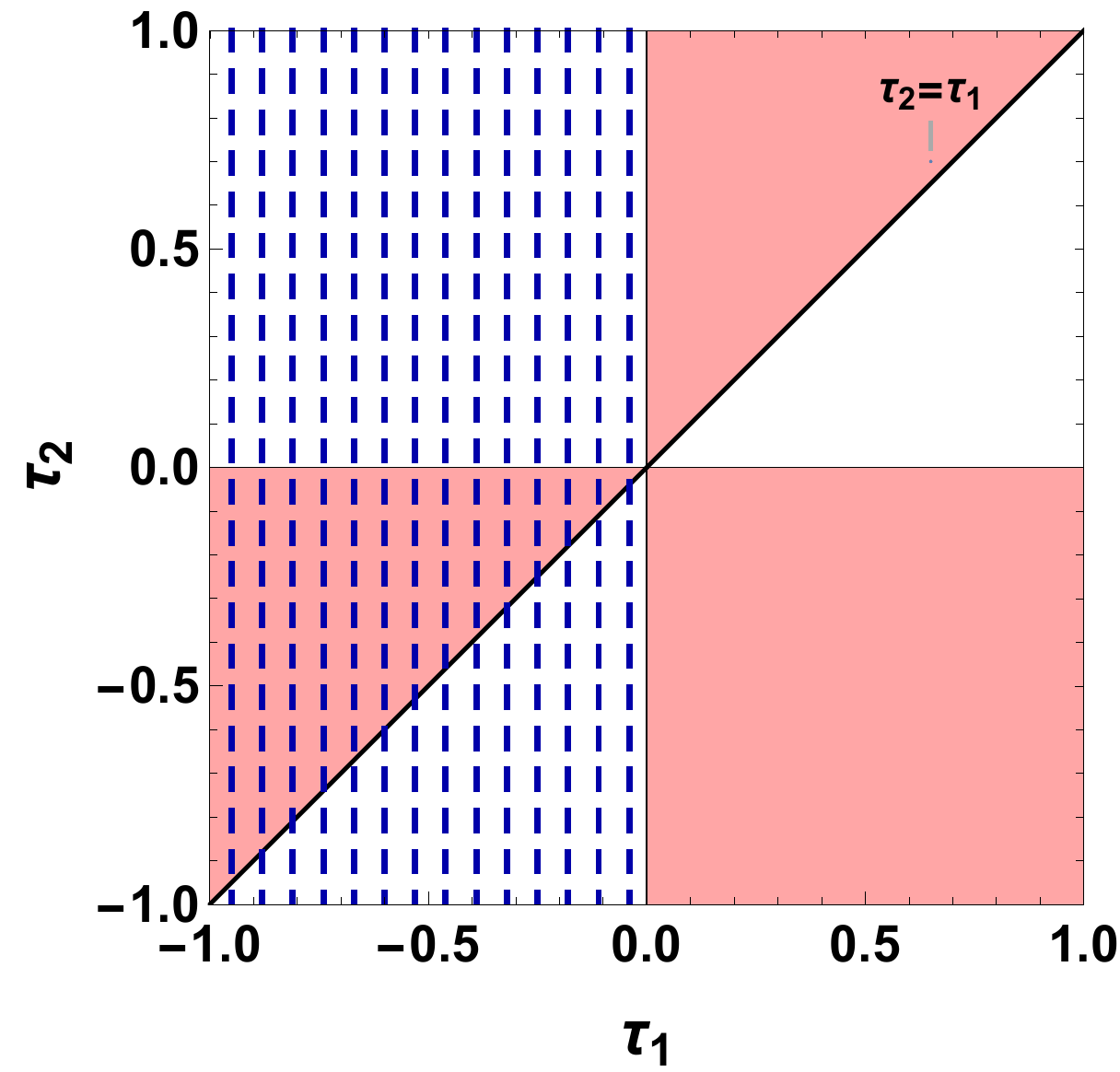}
		\includegraphics[width=0.49\linewidth]{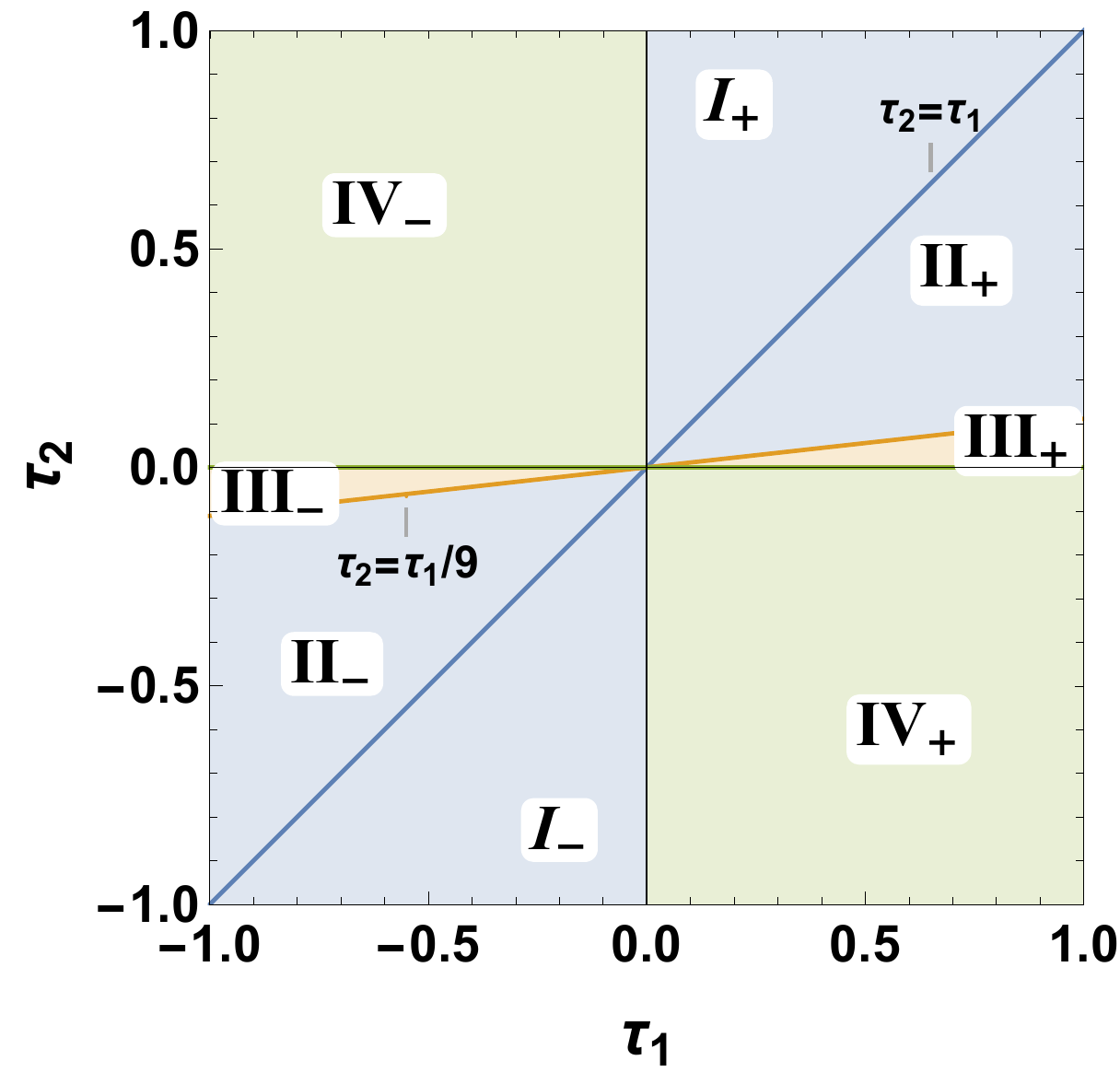}
	\end{center}
	\caption{Left: Parametric regions of the misbalance timescales $\tau_1$ and $\tau_2$ (\ref{eq:tau1}), (\ref{eq:tau2}) showing amplification of slow MA waves (red shading, imaginary parts of two out of three roots (\ref{roots_eq}) are negative regardless of the sign of the other root for all $n$) and entropy wave (blue dashed lines, imaginary part of one out of three roots (\ref{roots_eq}) is negative regardless of the sign of the other two roots for all $n$). {{The amplification/attenuation of slow MA and entropy waves corresponds to the regimes of their instability/stability, respectively.}}
		Right: Parametric regions of $\tau_1$ and $\tau_2$ showing different behaviour of slow MA and entropy waves, prescribed by the sign of the discriminant $\Delta$ (\ref{discriminant_eq}).
		The blue-shaded regions corresponds to $\Delta<0$ for all harmonics $n$ (two propagating slow MA modes and one non-propagating entropy mode, see \textit{Case 1} in Sec.~\ref{Subsec 3_1}). The green- and yellow-shaded regions show the regimes in which the discriminant $\Delta$ can be both positive and negative depending on the harmonic number $n$ (see Table \ref{tabcorDN}). In this case, all three modes become non-propagating for $\Delta>0$ in a certain range of harmonic numbers $n$ (see \textit{Case 2} in Sec.~\ref{Subsec 3_1}). The Roman numerals indicate specific regions of $\tau_{1,2}$ with different scenarios of the evolution of the entropy and slow MA waves, described in detail in Sec.~\ref{sec:param-reg} and demonstrated in Sec.~\ref{sec:spa-temp-evol}. We also recall that the blue ($\tau_2=\tau_1$) and yellow ($\tau_2 =\tau_1 / 9$) lines are equivalent to $\gamma_Q=\gamma$ and $\gamma_Q=\gamma / 9$, respectively, see Eqs.~(\ref{step_3})--(\ref{eq:tau2}).
		In both panels, the misbalance timescales $\tau_{1,2}$ are normalised to the isothermal acoustic travel time along the loop, $L/c_{S0}$ (see Sec.~\ref{sec:wave_eq}).
	}
	\label{Ampdamp}
\end{figure}

Thus, using Eqs.~(\ref{roots_eq})--(\ref{re_im_frq_ac_eq}) we {can write} conditions for the plasma modes to amplify/attenuate {{(be unstable/stable)}}, in terms of the characteristic misbalance times $\tau_1$ and $\tau_2$ (\ref{eq:tau1}), (\ref{eq:tau2}).
These conditions are visualised for the entropy and slow MA (both propagating and non-propagating) modes in the left-hand panel of Fig.~\ref{Ampdamp}. {In this work, we analyse the dependence of the  roots  $\omega_{1,2,3}$ (\ref{roots_eq}) on the characteristic times numerically, i.e.  we numerically search for the real and imaginary parts of $\omega_{1,2,3}$ for all possible values of $\tau_{1,2}$ (from $-\infty$ to $+\infty$).} 

The slow MA waves amplify {{(become unstable)}} if the imaginary part of two frequencies in the set of roots (\ref{roots_eq}) is negative regardless of the sign of the increment/decrement of the other root, thus associated with the entropy wave, for all harmonic numbers $n$. {As shown by Fig.~\ref{Ampdamp}, this condition is satisfied in three distinct regions of positive and negative $\tau_{1,2}$. These regions coincide with the isentropic instability condition for slow MA modes introduced by \citet{1965ApJ...142..531F} and written in terms of the characteristic times $\tau_{1,2}$ by \citet{2019PhPl...26h2113Z} and \citet{2019A&A...628A.133K} as
{{\begin{equation}\label{slow_ma_inst}
\frac{\tau_1-\tau_2}{\tau_1\tau_2}<0,
\end{equation}	}}in the limit of weak dispersion.} 

Similarly, the entropy mode is considered to amplify {{(become unstable)}}, if the imaginary part of one of three roots (\ref{roots_eq}) is negative regardless of the sign of the other two slow MA roots for all harmonic numbers $n$. This condition is satisfied for
{{\begin{equation}\label{entropy_inst}
\tau_1<0,
\end{equation}}}which corresponds to the isobaric instability \citep[see][]{1965ApJ...142..531F}. The same condition for the instability of the entropy mode was used by \citet{2020A&A...644A..33K}, in the context of stability of the solar corona and seismological constraining the coronal heating function.

\begin{table}[]
	\caption{Relationship between the discriminant sign  $\Delta \lessgtr 0$ and the harmonic number $n$ in different parametric regions of the characteristic times $\tau_1$ and $\tau_2$, shown in Fig.~\ref{Ampdamp}. Negative discriminant $\Delta<0$ indicates the regime of two propagating and one non-propagating modes (see \textit{Case 1} in Sec.~\ref{Subsec 3_1}). For positive discriminant $\Delta>0$, all modes become non-propagating (see \textit{Case 2} in Sec.~\ref{Subsec 3_1}). The critical harmonic numbers $n_{cr1}$ and $n_{cr2}$, determining the switch between those regimes, are given in Eq.~(\ref{crit_nums_eq}).}\label{tabcorDN}
	\tiny
	\begin{center}
		\begin{tabular}{|p{2.3cm}|p{1cm}|p{1cm}|p{1cm}|p{1cm}|p{1cm}|p{0cm}}
			\hline
			{\textbf{Regions  $I_{\pm}, II_{\pm}$}  \newline   $n_{cr1,2}$ are complex \newline (see blue in Fig.~\ref{Ampdamp}) }
			& \multicolumn{6}{m{9.0cm}|}{\raggedright $\Delta<0 $ for  all $n$}     
			\\ \hline
			
			{\textbf{Regions $III_{\pm} ~~~~~ $} \linebreak $n_{cr1,2}$ are real \newline (see yellow in Fig.~\ref{Ampdamp}) }
			& \multicolumn{2}{p{3.2cm}|}
			{\centering if   $ n_{cr1}>n_{cr2}\geq1:   $ \linebreak \linebreak
				\raggedright
				$\Delta<0 $,  $1\le n \le n_{cr2} $ \linebreak
				$\Delta>0 $,  $n_{cr2}+1 \le n \le n_{cr1} $ \linebreak
				$\Delta<0 $,  $n >  n_{cr1}  $ 			
			} 
			& \multicolumn{2}{p{2.6cm}|}
			{\raggedright if  $n_{cr1}\geq1,n_{cr2}=0:$  \linebreak  \linebreak
				$\Delta>0 $,  $1\le n \le n_{cr1} $	 \linebreak
				$\Delta<0 $,  $n >  n_{cr1}  $   
			} 
			& \multicolumn{2}{p{2.3cm}|}
			{\centering  if $n_{cr1} = n_{cr2} = 0:$  \linebreak  \linebreak
				$\Delta<0 $ for all $n$
			}
			\\ \hline
			
			{\textbf{Regions $IV_{\pm} ~~~~~ $} \linebreak  $n_{cr1}$ is real \newline   $n_{cr2}$ is complex  \newline (see green in Fig.~\ref{Ampdamp})   }
			& \multicolumn{3}{p{4.1cm}|}
			{\raggedright if $n_{cr1}\geq1:$  \linebreak \linebreak 
				$\Delta>0 $,  $1\le n \le n_{cr1}  $  \linebreak
				$\Delta<0 $,  $n >  n_{cr1}  $ 
			}
			& \multicolumn{3}{p{3.6cm}|}
			{\raggedright if $  n_{cr1}=0:$  \linebreak  \linebreak
				$\Delta<0 $ for all $n$ }             \\ \hline
		\end{tabular}
	\end{center}
\end{table}


In order to determine where in the spectrum (i.e. a specific range of the perturbation harmonic numbers $n$ and the corresponding wavenumbers $k$) switching between \textit{Cases 1} and \textit{2} takes place, we solve $\Delta = 0 $ (\ref{discriminant_eq}) with respect to $ k^2$. This gives us three critical wavenumbers, $k_{cr0}$, $k_{cr1}$, and $k_{cr2}$. 
{One of these critical  wavenumbers is equal to 0 (i.e. $ k_{cr0}=0$ and $n_{cr0}=0$ ), thus corresponding to the non-oscillatory background (see Sec.~\ref{Subsec 3_1}). }
The other two critical  wavenumbers are
\begin{equation}\label{crit_eigen_eq1}
	k_{cr1,2}=\sqrt{\frac{\nu_2^2}{8\gamma^3} \left[\left(\gamma^2+18\gamma\gamma_Q -27\gamma_Q^2 \right) \pm \left(\gamma_Q-\gamma\right)^\frac{1}{2} \left(9\gamma_Q-\gamma\right)^\frac{3}{2} \right]}.
\end{equation}
These eigenvalues, in turn, give the critical harmonic numbers at which the change of the discriminant sign happens,
\begin{equation}\label{crit_nums_eq}
	n_{cr1}=\mathrm{floor}\left(\frac{k_{cr1}}{\pi}l\right),~~n_{cr2}=\mathrm{floor}\left(\frac{k_{cr2}}{\pi}l\right).
\end{equation}

It is clearly seen that the critical harmonic numbers $n_{cr1}$,  $n_{cr2}$ depend on the absolute value and ratio of the characteristic timescales $\tau_1$ and $\tau_2$. {These numbers can be either real or complex defining the boundaries of the non-propagating harmonic range (if $n_{cr1},n_{cr2}$ are real, then $n_{cr1}>n_{cr2}$). }  Thus, the parametric regions of $\tau_{1,2}$ with different sign of the discriminant $\Delta$ (\ref{discriminant_eq}) across the perturbation spectrum are shown in the right-hand panel of Fig.~\ref{Ampdamp}. The blue-shaded regions in Fig.~\ref{Ampdamp} indicate the regime with $\Delta<0$ for all harmonic numbers $n$. It means that \textit{Case 1} holds true for any harmonic in the spectrum, i.e. all slow MA harmonics propagate (see also Table~\ref{tabcorDN}). 
The yellow shading in Fig.~\ref{Ampdamp} indicates the region where the discriminant may be either positive or negative. This implies that in some range of the harmonic numbers $n$ the slow modes become non-propagating (\textit{Case 2}). This range is determined by the critical harmonic numbers $n_{cr1}$ and $n_{cr2}$ (\ref{crit_nums_eq}). Depending on the absolute value of the characteristic times $\tau_{1,2}$, it may be located at high or low harmonic numbers $n$ and include numerous harmonics (see Table~\ref{tabcorDN}). For a particular case with $n_{cr2}=0$, this range starts from the fundamental harmonic $n=1$ which thus becomes non-propagating. The green shading in Fig.~\ref{Ampdamp} indicates the region where the discriminant may be either positive or negative too. However, in this case the range of non-propagating slow MA  harmonics (\textit{Case 2}) may only start from the fundamental harmonic $n=1$ ($n_{cr2}$ is complex) and is limited by $n_{cr1}$. For $n_{cr1}=n_{cr2}=0$ in the yellow-shaded regions and for $n_{cr1}=0$ in the green-shaded regions in Fig.~\ref{Ampdamp}, \textit{Case 2} degenerates to \textit{Case 1}, i.e. all slow MA harmonics propagate.

In summary, the solution of Eq. (\ref{temporal_eq}) gives us temporal dependence $\psi\left(t\right)$ of the full solution $\rho(z,t)$ of the wave equation (\ref{lin_evol_eq}). It may have two different forms, referred to as \textit{Cases 1} and \textit{2} in this section. The difference is caused by the fact that slow MA harmonics may become non-propagating in a certain range of the perturbation spectrum depending on the characteristic thermal misbalance timescales $\tau_{1,2}$. The range of non-propagating harmonics may be located at different parts of the spectrum (i.e. at high or low harmonic numbers), and is determined by the critical harmonic numbers $n_{cr1}$ and  $n_{cr2}$ (\ref{crit_nums_eq}). The parametric regions of the characteristic times $\tau_{1,2}$ where the change of the solution form takes place are demonstrated in Fig.~\ref{Ampdamp} and Table~\ref{tabcorDN}.
In addition to the effect of non-propagating slow MA harmonics, thermal misbalance leads to the amplification/attenuation of slow MA and entropy modes (see the left-hand panel of Fig. \ref{Ampdamp}). {{Thus, a combination of the left-hand and right-hand panels in Fig.~\ref{Ampdamp} should be used for determining values of $\tau_{1,2}$ which allow for (a) stable/unstable behaviour of slow MA and entropy waves, and (b) propagation/non-propagation of slow MA harmonics.}}
 In Sec.~\ref{sec:param-reg}, we discuss the dependence of the slow MA speed and increments/decrements of slow MA and entropy modes on the harmonic number (i.e. an effective dispersion and frequency-dependent damping/amplification of slow MA and entropy waves, respectively), also caused by the phenomenon of thermal misbalance. A synergy of these effects leads to a number of different scenarios for the initial perturbation to evolve, which are outlined in Sec.~\ref{sec:param-reg} and demonstrated in Sec.~\ref{sec:spa-temp-evol}.

\subsection{Parametric regions with different behaviour of the individual Fourier harmonics}
\label{sec:param-reg}

\begin{figure*}
	\begin{center}
		\includegraphics[width=1.0\linewidth]{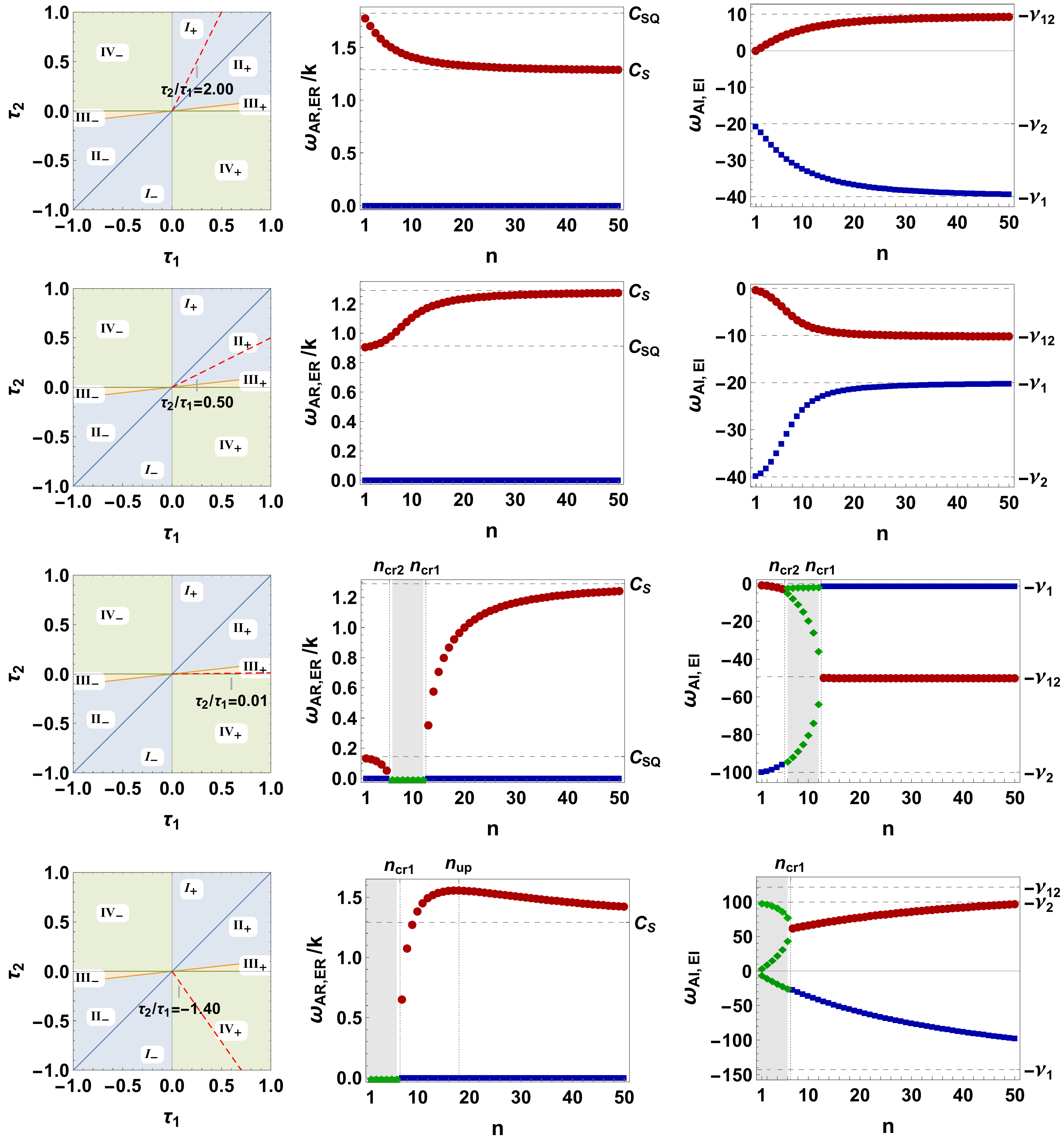}
	\end{center}
	\caption{{Left column} indicates the regions of the thermal misbalance timescales $\tau_{1,2}$, in which the dispersive properties of slow MA and entropy waves are qualitatively different (see also Fig.~\ref{Ampdamp}), with $\tau_2/\tau_1=2$ in region $I_+$ (the top row), $\tau_2/\tau_1=0.5$ in region $II_+$ (the second row), $\tau_2/\tau_1=0.01$ in region $III_+$ (the third row), and $\tau_2/\tau_1=-1.4$ in region $IV_+$ (the bottom row).		
	Middle and right columns show the dependence of the phase speed $\omega_{AR,ER}/{k}$ and increment/decrement $\omega_{AI,EI}$ of the slow MA and entropy modes on the harmonic number $n$, respectively.
	{The dependencies are calculated using Eqs. (\ref{im_frq_ent_eq}), (\ref{re_im_frq_ac_eq}) for harmonics $n$ satisfying  \textit{Case 1}, and general expressions (\ref{roots_eq}) for harmonics $n$ satisfying  \textit{Case 2}. }		
	The red symbols correspond to the harmonics of two propagating slow MA waves. The blue symbols correspond to the entropy mode. {The green symbols correspond to the harmonics satisfying  \textit{Case 2}, for which all three modes become non-propagating.} The ranges of $n$, in which those non-propagating slow MA harmonics appear, are determined by the critical harmonic numbers $n_{cr1,2}$ (\ref{crit_nums_eq}) and shown by {grey shading.}  The harmonic number $n_{up}$ shows the change of the sign of the gradient of the dependence $\omega_{AR}(n)$ in region $IV_+$, and is obtained from the condition $\partial \omega_{AR} / \partial n = 0$. The limiting values of the slow MA phase speed, $c_{S}$ and $c_{SQ}$ are given in Eq.~(\ref{step_3}), and normalised to the isothermal sound speed $c_{S0}$. The limiting values of the slow MA and entropy increments/decrements, $\nu_{1,2}$ and $\nu_{12}$ are given by Eq.~(\ref{chr_frq_2}).
		}
	\label{DispTab}
\end{figure*}

A combination of the left panel (regions of damping/amplification of slow MA and entropy modes) and right panel (regions of propagating/non-propagating slow MA harmonics) in Fig.~\ref{Ampdamp} allows us to distinguish parametric regions of $\tau_{1,2}$ with qualitatively different spatio-temporal behaviour of slow MA and entropy waves (the Roman numerals in Fig.~\ref{Ampdamp}). Quantitatively, the difference in the wave behaviour is caused by the dependence of the phase speed of slow MA waves (the phase speed of the entropy wave is always zero, $\omega_{ER}/k\equiv0$) and increment/decrement of slow MA and entropy waves on the harmonic number $n$ (the wavenumber $k$), which are different for different combinations of the misbalance parameters $\tau_{1,2}$. To illustrate this, we calculated  dependencies of the wave increment/decrement and phase speed on the harmonic number $n$ for slow MA and entropy waves in some of those regions (see Fig.~\ref{DispTab}).
{We used Eqs.~(\ref{im_frq_ent_eq}) and (\ref{re_im_frq_ac_eq}) for the harmonics $n$ satisfying  \textit{Case 1}, and general expressions (\ref{roots_eq}) for the harmonics $n$ satisfying \textit{Case 2}. }

{As it was shown in the previous works \citep[e.g.][]{2019PhPl...26h2113Z,2019A&A...628A.133K}, the thermal misbalance has a weak impact on the dispersion properties of waves in the short-wavelength limit ($n \to \infty$) and strongly affects them in the long-wavelength limit ($n \to 0$).
Assuming that harmonics in both of these limits satisfy \textit{Case 1}, we can write  the increments/decrements of entropy and slow MA waves as
\begin{align}\label{ent_av_inc_lim}
	&\lim_{n \to 0} \omega_{EI}=-\nu_2, & &\lim_{n \to \infty} \omega_{EI}=-\nu_1, \\
	& \lim_{n \to 0} \omega_{AI}=0, & &\lim_{n \to \infty} \omega_{AI}=-\nu_{12},\nonumber
\end{align}
with the parameters $\nu_1$, $\nu_2$, and $\nu_{12}$ determined by Eq.~(\ref{chr_frq_2}). Equations~(\ref{ent_av_inc_lim}) have no contradiction with the results shown in Fig.~\ref{Ampdamp}, stating that growth/decay of the entropy wave is defined by the sign of the characteristic time $\nu_1$ $(\tau_1)$ for all $n$. For long-wavelength harmonics  ($n \to 0$), \textit{Case 1} occurs only if the characteristic times $\nu_{1,2}$ $(\tau_{1,2})$  are of the same sign (see the first to third rows in Fig.~\ref{DispTab}). Thus, the negative/positive  $\nu_2$ $(\tau_2)$  is equivalent to the negative/positive  $\nu_1$ $(\tau_1)$ in this case. }


Due to the effect of non-propagation, the slow MA phase speed may have two values in the long-wavelength (strong misbalance) limit,
\begin{align}\label{ac_ph_n0_lim}
	&\lim_{n \to 0} \frac{\omega_{AR}}{k}=c_{SQ},&&\mbox{for propagating slow MA waves, \textit{Case 1}}\\
	&\lim_{n \to 0} \frac{\omega_{AR}}{k}= 0,&&\mbox{for non-propagating slow MA waves, \textit{Case 2}}\nonumber
\end{align}
as shown by the first to third rows and the fourth row in Fig.~\ref{DispTab}, respectively.
In the short-wavelength limit (weak misbalance), the effect of non-propagation never takes place and slow MA phase speed is not affected by the misbalance process, so that
\begin{equation}\label{ac__ph_ninf_lim}
	\lim_{n \to \infty} \frac{\omega_{AR}}{k} = c_{S}.
\end{equation}

\textit{Entropy mode in regions  $I_{\pm}$--$IV_{\pm}$.} As shown by Fig.~\ref{DispTab}, in all {regions  $I_{+}$} to $IV_{+}$ the entropy mode decays ($\omega_{EI}<0$). Being excited simultaneously with slow MA modes, the entropy mode contains a part of the initial perturbation energy. Thus, after the slow MA waves run away from the site of the initial perturbation, the essentially non-propagating entropy mode forms a localised density and temperature disturbances which decay with time. However, the decay rate of the individual entropy harmonics is different in those regions. Namely, in {regions $I_{+}$} and $IV_{+}$ ($II_{+}$ and $III_{+}$) the shorter- (longer-) wavelength harmonics decay more efficiently, respectively. The situation is symmetrically opposite in {regions  $I_{-}$} to $IV_{-}$, where $\omega_{EI}>0$ and the entropy mode grows (not shown in Fig.~\ref{DispTab}).


\textit{Slow MA modes in region $I_{+}$.} According to the top row in Fig.~\ref{DispTab}, slow MA waves propagate and amplify in {region $I_{+}$} ($\omega_{AI}>0$), with higher increment at shorter-wavelength harmonics. In this region also, the phase speed of slow MA modes experiences a negative dispersion, i.e. longer-wavelength harmonics propagate faster than those with shorter-wavelength. A combination of these effects leads to the formation of propagating quasi-periodic patterns, which could be referred to as slow MA wave trains. The linear stage of their formation (with the relative amplitude of the plasma density perturbation $\rho \ll 1$) from the initial Gaussian pulse has been discussed in detail in \citet{2019PhPl...26h2113Z} in terms of linear Eq.~(\ref{lin_evol_eq}). In a weakly nonlinear regime with $\rho \lessapprox 1$, the growing slow MA wave develops into self-sustained shock pulses, for description of which a nonlinear evolutionary equation is required \citep[see e.g.][]{2020PhRvE.101d3204Z,2011Ap&SS.334...35M}. The strong amplification may also lead to highly nonlinear variations with $\rho \gtrapprox 1$. In this case, parameters of the nonlinear slow MA shock structures can be {found using the solution of} the full set of hydrodynamic equations \citep{2020TePhL..46..637M}.  

\textit{Slow MA modes in region $II_{+}$.} The second row of Fig. \ref{DispTab} shows properties of slow MA modes in {region $II_{+}$}. In this case, slow MA waves propagate and decay ($\omega_{AI}<0$) with higher decrement at shorter-wavelength harmonics.
The phase speed of slow MA modes has a positive dispersion, with shorter-wavelength harmonics overtaking.
In this case, the effect of dispersion and frequency-dependent damping causes the initial Gaussian pulse to become asymmetric, broaden, and decrease in its amplitude with time \citep[see][for details]{2019PhPl...26h2113Z}. As both slow MA waves decay in this case, the description of nonlinear effects is required only in case of a nonlinear initial perturbation.  

\textit{Slow MA modes in region $III_{+}$.} In this region of the misbalance parameters $\tau_{1,2}$ (see the third row of Fig.~\ref{DispTab}), the range of non-propagating slow MA harmonics may appear in the spectrum (see \textit{Case 2} in Sec.~\ref{Subsec 3_1}), that is determined by the critical wavenumbers $n_{cr1}$ and $n_{cr2}$ (\ref{crit_nums_eq}).
{Both the propagating slow MA waves with $\omega_{AI}<0$ and all non-propagating harmonics with purely imaginary $\omega_{1,2,3}<0$ decay in this region.}
The dispersion of the phase speed can be either positive or negative in this region. For harmonics $n<n_{cr2}$, the dispersion is negative and $|\omega_{AI}|$ is lower (weaker damping); for harmonics $n>n_{cr1}$, it becomes positive and $|\omega_{AI}|$ is higher (stronger damping). Potentially, formation of quasi-periodic slow MA structures from a broadband initial perturbation due to the negative dispersion and weak damping in the long-wavelength band is possible. {However, the effect is to be less pronounced than in region $I_{+}$ due to a small number of harmonics $n<n_{cr2}$, and relatively low variation of their phase speed between $c_{SQ}$ and 0. Indeed, in this region, the maximum value of $c_{SQ}$ prescribed by $ \gamma_Q = {\gamma\tau_2}/{\tau_1}$ is $c_{S}/3$ that is associated with the upper boundary of region $III_{+}$, $0<\tau_2/\tau_1<1/9$.}

\textit{Slow MA modes in region $IV_{+}$.} In this region (see  the bottom row in Fig.~\ref{DispTab}), the range of non-propagating slow MA harmonics may also appear in the spectrum. However, in contrast to region $III_{+}$, slow MA waves grow with $\omega_{AI}>0$. Similarly to region  $III_{+}$, the non-propagating slow MA harmonics may have both higher and lower increments relative to the propagating harmonics.  
The phase speed of slow MA waves is a non-monotonic function of the harmonic number $n$, providing either positive or negative dispersion (see the middle panel in the bottom row of Fig.~\ref{DispTab}). The range of propagating harmonics with positive dispersion has a finite number of harmonics. {The lower limit of this range is determined by the critical number $n_{cr1}$ (\ref{crit_nums_eq}). The upper limit, $n_{up}$  can be found by solving the equation $\partial \omega_{AR} / \partial n = 0$ (\ref{re_im_frq_ac_eq}}). The range of harmonics with negative dispersion, in turn, has no upper limit. A combination of the negative dispersion and amplification again may lead to formation of  quasi-periodic slow MA patterns. The effect will be stronger pronounced for low values of $n_{up}$, so that the majority of slow MA harmonics will have the negative dispersion. The dispersion properties of slow MA waves in this case will be qualitatively similar to those in region $I_+$.
The amplifying non-propagating slow MA harmonics in region $IV_{+}$, which occur for the harmonic numbers $1<n<n_{cr1}$, would develop into long-wavelength density disturbances, similarly to those caused by the entropy mode, thus making the plasma essentially non-uniform along the field. 

\textit{In regions $I_{-}$ to $IV_{-}$}, the properties of slow MA waves are symmetrically opposite to those in regions $I_{+}$ to $IV_{+}$ described above. Hence, these regions are not shown in Fig.~\ref{DispTab}.

Thus, the dispersive properties of slow MA and entropy waves are shown to strongly depend on the characteristic timescales $\tau_1$ and $\tau_2$ of the thermal misbalance process, that may lead to dramatically different scenarios for the evolution of the initial broadband perturbation of the coronal plasma. For illustration, in Sec.~\ref{sec:applications} we show how entropy and slow MA modes excited simultaneously share the initial perturbation energy and evolve in the linear regime in regions $I_{\pm}$ and $II_{\pm}$, using the exact full solution $\rho(z,t)$ presented in Sec.~\ref{Subsec 4_1}.

\section{Exact solution}
\label{Subsec 4_1}

\begin{table}
	\caption{Exact solution of Eq.~(\ref{lin_evol_eq}) in different regions of the characteristic thermal misbalance timescales $\tau_{1,2}$ (see Fig.~\ref{Ampdamp}), for $ n_{cr1}>n_{cr2}>1$ (\ref{crit_nums_eq}). The functions $\rho_{n\Delta-}\left(z,t\right)$, $\rho_{n\Delta+}\left(z,t\right)$, and $\rho_{0}\left(z,t\right)$ are given in Eqs.~(\ref{rhon_delt_min_eq}), (\ref{rhon_delt_plus_eq}), and (\ref{rho0_delt_0_eq}), respectively.}\label{ExactSolTab}
		\begin{tabular}{|c|c|}
			\hline
			Regions & Solution \\
			\hline
			 $I_\pm, II_\pm$ & $	\rho\left(z,t\right) = 	\rho_{0}\left(z,t\right) + \sum\limits_{n=1}^{\infty} \rho_{n\Delta-}\left(z,t\right) $ \\
			\hline
			 $III_\pm$ &  $	\rho\left(z,t\right) = 	\rho_{0}\left(z,t\right) + \sum\limits_{n=1}^{ n_{cr2} } \rho_{n\Delta-}\left(z,t\right) + $		 \\
				 &  $ \sum\limits_{n =  n_{cr2}+1 }^{ n_{cr1} } \rho_{n\Delta+}\left(z,t\right)  +	\sum\limits_{n = n_{cr1} +1}^{\infty} \rho_{n\Delta-}\left(z,t\right)   $ \\
			\hline
			 $IV_\pm$ &  $	\rho\left(z,t\right) = 	\rho_{0}\left(z,t\right) + \sum\limits_{n=1}^{ n_{cr1} } \rho_{n\Delta+}\left(z,t\right) +
			\sum\limits_{n =   n_{cr1}+1}^{\infty} \rho_{n\Delta-}\left(z,t\right)  $ \\
			\hline
		\end{tabular}
\end{table} 
In Sec.~\ref{sec:parametric-analysis}, we have discussed the spatial $\varphi\left(z\right)$ and temporal $\psi\left(t\right)$ dependencies of the full solution $\rho\left(z,t \right) = \varphi\left(z\right) \psi\left(t\right)$, described by Eqs.~(\ref{spatial_eq}) and (\ref{temporal_eq}), respectively. Using solutions to these equations, in this section we present the full exact solution of the governing evolutionary Eq.~(\ref{lin_evol_eq}) for the density perturbation $\rho\left(z,t \right)$ caused by a superposition of slow MA and entropy waves with harmonic number $n$. There are three possible cases:

\begin{enumerate} 
	\item The solution for the $n$-th harmonic with  $\Delta<0$ (\textit{Case 1}: two oppositely propagating slow MA modes and one non-propagating entropy mode),
	\begin{eqnarray}\label{rhon_delt_min_eq}
		&\rho_{n\Delta-}\left(z,t\right) = C_{1n} e^{\omega_{EI}t} \cos \left( k z\right) + \\
		&C_{0n} e^{\omega_{AI}t}   \left[\cos \left(\omega_{AR} t + k z  - \phi_n \right)+ \cos \left(\omega_{AR} t - k z  - \phi_n \right) \right],  \nonumber
	\end{eqnarray}
	where
	\begin{equation}\label{not_rhon_delt_min}
		C_{0n}=\frac{\sqrt{C_{2n}^2+C_{3n}^2}}{2},~~\phi_n = \arctan \left(\frac{C_{3n}}{C_{2n}}\right).
	\end{equation}
	The constants $	C_{1n}$, $C_{2n}$, and  $C_{3n}$ can be obtained by solving the following set of linear equations,
	\begin{equation}\label{const_mat_delt_min}
		\begin{pmatrix} 1 & 1 & 0 \\ \omega_{EI} & -\omega_{AI} & \omega_{AR} \\ \omega_{EI}^2 & \left(\omega_{AI}^2-\omega_{AR}^2\right) & -2\omega_{AR}\omega_{AI} \\  \end{pmatrix}   		\begin{pmatrix} C_{1n}  \\ C_{2n} \\ C_{3n}  \end{pmatrix}  	 =		\begin{pmatrix} I_{1n}   \\  I_{2n} \\ I_{3n}  \end{pmatrix} .
	\end{equation}
	The integrals $I_{1n}$, $I_{2n}$, and $I_{3n}$  are prescribed by the initial perturbation $\rho_{in}(z,0)$ and the derivatives $\at{(\partial \rho(z,t) /  \partial t)}{t=0}$, and $ \at{(\partial^2 \rho(z,t) /  \partial t^2)}{t=0}$ as
	\begin{align}
		&I_{1n}  = \frac{2}{l} \int_0^l \rho_{in}(z,0) \cos \left( k z\right) dz,\label{init_integral_1}\\
		&I_{2n}  = \frac{2}{l} \int_0^l \at{\frac{\partial \rho(z,t)}{\partial t}}{t=0}  \cos \left( k z\right) dz,\label{init_integral_2}\\
		&I_{3n} = \frac{2}{l} \int_0^l \at{\frac{\partial^2 \rho(z,t)}{\partial t^2}}{t=0} \cos \left( k z\right) dz.\label{init_integral_3}
	\end{align}

	\item The solution for the $n$-th harmonic with  $\Delta>0$ (\textit{Case 2}: all three modes non-propagating),
	\begin{equation}\label{rhon_delt_plus_eq}
		\rho_{n\Delta+}\left(z,t\right) = \left( C_{1n} e^{-i\omega_1 t} + C_{2n} e^{-i\omega_2 t} +C_{3n} e^{-i\omega_3 t} \right)  \cos \left( k z\right).
	\end{equation}
	The constants $	C_{1n}$, $C_{2n}$, and  $C_{3n}$ can be obtained by solving the following set of linear equations,
	\begin{equation}\label{const_mat_delt_plus}
		\begin{pmatrix} 1 & 1 & 1 \\ \omega_1 & \omega_2 & \omega_3 \\ \omega_1^2 & \omega_2^2 & \omega_3^2 \\  \end{pmatrix} 		\begin{pmatrix} C_{1n}  \\ C_{2n} \\ C_{3n}  \end{pmatrix}  	 =		\begin{pmatrix} I_{1n}   \\  I_{2n}  \\ I_{3n}   \end{pmatrix},
	\end{equation}
	with the integrals $I_{1n}$, $I_{2n}$, and $I_{3n}$ determined by Eqs.~(\ref{init_integral_1})--(\ref{init_integral_3}).
	
	\item A non-oscillating and non-propagating background value with $ \Delta=0 $ and $ n = 0 $ is 
	\begin{equation}\label{rho0_delt_0_eq}
		\rho_{0}\left(z,t\right) =  C_{10} e^{-\nu_2 t} +  C_{20} t + C_{30}.
	\end{equation}
	The constants  $C_{10}$, $C_{20}$, and  $C_{30}$  can be obtained by solving the following set of linear equations,
	\begin{equation}\label{const_mat_delt_0}
		\begin{pmatrix} 1 & 0 & 1 \\ -\nu_2 & 1 & 0 \\ \nu_2^2 & 0 & 0 \\  \end{pmatrix} 		\begin{pmatrix} C_{10}  \\ C_{20} \\ C_{30}  \end{pmatrix}  	 =		\begin{pmatrix} I_{10}   \\  I_{20} \\ I_{30} \end{pmatrix},
	\end{equation}
	where
	\begin{align}
		&I_{10}  = \frac{1}{l} \int_0^l \rho_{in}(z,0) dz\label{init_integral_10},\\
		&I_{20}  = \frac{1}{l} \int_0^l \at{\frac{\partial \rho(z,t)}{\partial t}}{t=0}  dz\label{init_integral_20},\\
		&I_{30}  = \frac{1}{l} \int_0^l\at{\frac{\partial^2 \rho(z,t)}{\partial t^2}}{t=0}   dz\label{init_integral_30}.
	\end{align}
	
\end{enumerate}

Using Eqs.~(\ref{rhon_delt_min_eq}), (\ref{rhon_delt_plus_eq}), and (\ref{rho0_delt_0_eq}), we can construct the exact solution of Eq.~(\ref{lin_evol_eq}) by the superposition principle. {In other words, to obtain the exact solution we sum up  the solutions for all harmonics from $n=1$ to infinity, using  Eqs.~(\ref{rhon_delt_min_eq}) and (\ref{rhon_delt_plus_eq}) for harmonics satisfying $\Delta<0$ and $\Delta>0$, respectively, and add Eq.~(\ref{rho0_delt_0_eq}) for a non-oscillating background ($n=0$).} The series describing the exact solution for different regions of the characteristic thermal misbalance timescales $\tau_{1,2}$  (see Fig.~\ref{Ampdamp}) are presented in Table~\ref{ExactSolTab}. The obtained exact solution gives us plenty of possibilities for the analysis of the perturbation evolution.

\section{Applications of the exact solution}
\label{sec:applications}

The exact solution shown by Table~\ref{ExactSolTab}, in particular, allows us to study the initial (linear) stage of the perturbation evolution in any regions of the characteristic misbalance times  $\tau_1$ and $\tau_2$. Moreover, it allows us to study the evolution of slow MA and entropy waves separately, {attributing the first and second terms on the RHS of Eq.~(\ref{rhon_delt_min_eq}) to the entropy and slow MA mode, respectively. As it was discussed in Sec.~\ref{sec:parametric-analysis}, this distinct attribution is only possible for \textit{Case 1} with the discriminant $\Delta < 0$ (\ref{discriminant_eq}), while in \textit{Case 2} with $\Delta > 0$ and solution (\ref{rhon_delt_plus_eq}) it is less obvious because of the mixed properties of those waves. Hence, in this section, for illustration we apply the obtained exact solution to the evolution of the initial localised perturbation in parametric regions $I_{\pm}$ and $II_{\pm}$ (see Fig.~\ref{Ampdamp}), throughout of which \textit{Case 1} holds true, i.e. all slow MA harmonics propagate and all entropy harmonics do not propagate. }

\subsection{Spatio-temporal evolution of entropy and slow MA modes } \label{Subsec 4_2}
\label{sec:spa-temp-evol}

The specific form of the initial perturbation is determined by its functional dependence on the $z$-coordinate and what plasma parameters are perturbed.
This in turn determines the functions $ \rho_{in}(z,0)$, $\at{(\partial \rho(z,t) /  \partial t)}{t=0}$, and $ \at{(\partial^2 \rho(z,t) /  \partial t^2)}{t=0}$, which affect the partition of the initial perturbation energy between slow MA and entropy waves, according to Eq.~(\ref{const_mat_delt_min}). From the set of governing MHD equations written for the perturbations along the field in a zero-$\beta$ plasma \citep[see e.g. Eqs.~(1)--(4) in][]{2019PhPl...26h2113Z}, the derivatives ${\partial \rho(z,t) /  \partial t}$ and ${\partial^2 \rho(z,t) /  \partial t^2}$ (and hence their values at $t=0$) are connected with the perturbations of the other plasma parameters as 
\begin{align}
& \frac{\partial \rho(z,t)}{\partial t} = -{\frac{\partial V_{z}(z,t)}{\partial z}},\label{eq:drho/dt}\\
&\frac{\partial^2 \rho(z,t)}{\partial t^2}  ={\frac{\partial^2 P(z,t)}{\partial^2 z}}.\label{eq:d2rho/dt2}
\end{align}

In this work, we consider the initial Gaussian pulse perturbing the plasma density, pressure, and temperature, 
\begin{align}\label{init_gaussian} 
	&\rho_{in}\left(z,0\right)= A_{\rho} \exp\left[-\left(z-z_0\right)^2 / w\right],\nonumber \\ 
	&P_{in}\left(z,0\right)= A_{P} \exp\left[-\left(z-z_0\right)^2 / w\right],\\ 
	&T_{in}\left(z,0\right) = P_{in}\left(z,0\right) - \rho_{in}\left(z,0\right), \nonumber\\
	&V_{z,in}\left(z,0\right) = 0.  \nonumber
\end{align}
Here, the initial perturbation of the plasma velocity $V_{z,in}$ is taken to be zero, implying the absence of the injected plasma flows. All the initial perturbations (\ref{init_gaussian}) are normalised to the equilibrium values $\rho_0$, $ \rho_0 c_{S0}^2$, $ T_0$, and $ c_{S0} $, respectively; $A_{\rho}$ and $A_P$ are magnitudes of  the density and pressure variations; $w$ and $z_0$ are the effective width and position of the perturbing pulse, respectively. 
We assume here that $A_{\rho}=A_{P} $ which implies that the initial perturbation (\ref{init_gaussian}) is of an isothermal type (i.e. $T_{in}=0$). This assumption is justified, for example, for perturbations by impulsive heating events in the solar corona, in which the rapid increase in the plasma temperature strengthens the efficiency of the parallel thermal conduction that tends to smooth the temperature perturbation out, on the timescale much shorter than the timescale of the waves excited \citep[see e.g. a series of works by][]{2016ApJ...826L..20R, 2018ApJ...856...51R, 2019ApJ...884..131R}. In addition, it can be shown that for the initial perturbation (\ref{init_gaussian}), the coefficients $C_{10}$ and $C_{20}$ in Eq.~(\ref{rho0_delt_0_eq}) are equal to zero, so that the background value $\rho_0$ does not vary in time and, hence, does not affect the damping or amplification of entropy and slow MA modes.

\begin{figure}
	\begin{center}
		\includegraphics[width=\linewidth]{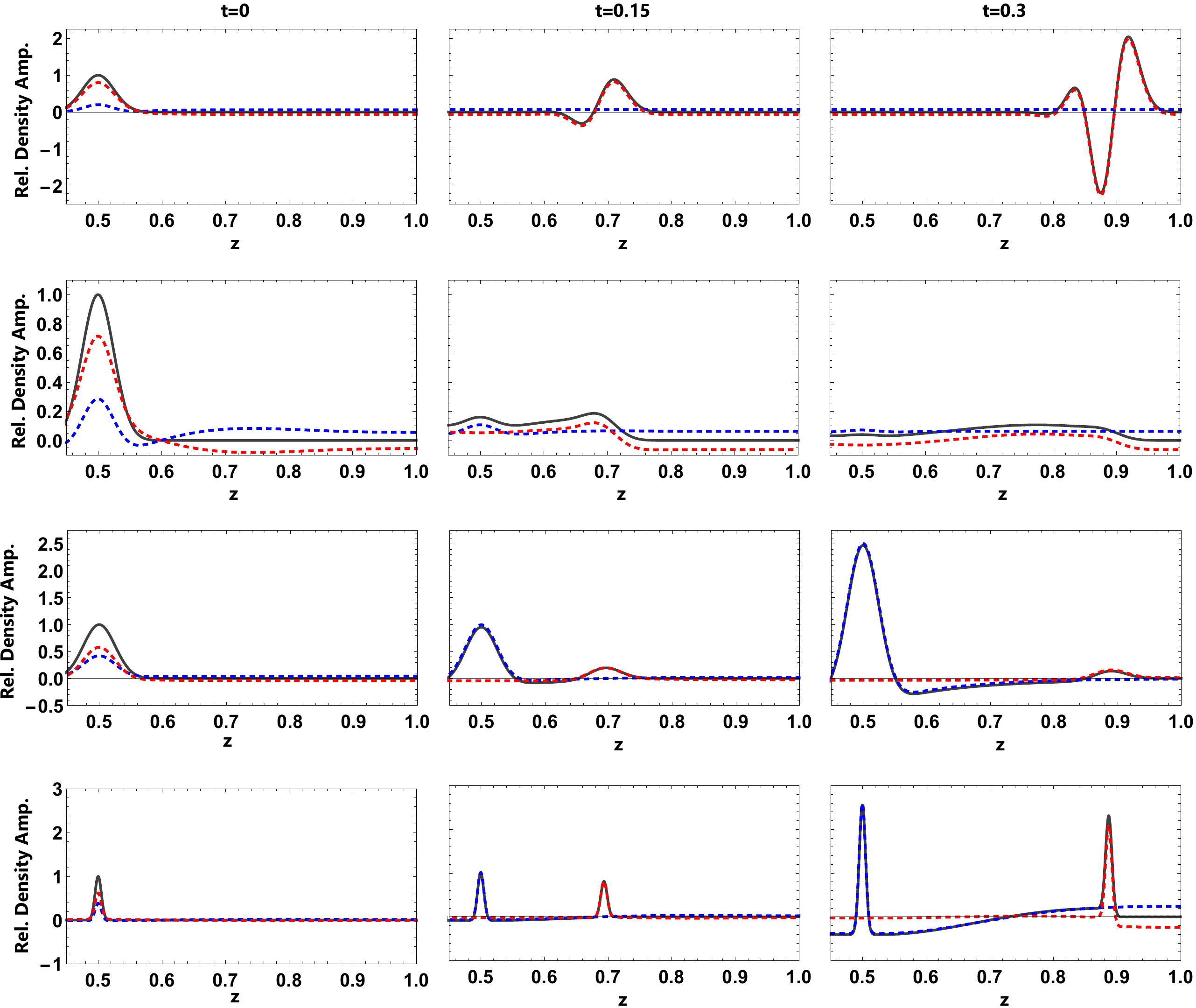}
	\end{center}
	\caption{{{Evolution of the initial plasma density perturbation of a Gaussian shape (\ref{init_gaussian}) situated at $z_0=l/2$, in the parametric regions
	$I_+$ with  $\tau_2 = 0.016$ and $\tau_1 = 0.01 $ (top row);
	$II_+$ with $\tau_2 = 0.04$ and $\tau_1 = 0.1 $ (second row);
	$I_-$ with $\tau_2 = -0.4$ and $\tau_1 = -0.15 $ (third row);
	and $II_-$ with $\tau_2 = -0.05$ and $\tau_1 = -0.15 $ (bottom row)
	of the characteristic misbalance timescales $\tau_{1,2}$ (see Fig.~\ref{Ampdamp}). The exact analytical solution describing the wave evolution in those regions is given in Table~\ref{ExactSolTab}. Left, middle and right columns indicate the solutions at $t=0$, $t=0.15$, and  $t=0.3$ of the computational time, respectively.  	The red and blue dashed lines correspond to one slow MA mode and one entropy mode, respectively. The solid black line corresponds to the full solution (sum of solutions for one entropy and two slow MA modes). }} 	
	The relative density amplitude on the vertical axis is shown in the units of the initial density pulse amplitude $A_{\rho}$, shared between one entropy wave and two slow MA waves.
	The horizontal axis is normalised to the characteristic spatial scale of the medium (for example, the loop length, $L$). 
	{The background value $\rho_0(z,t)$ (see Table~\ref{ExactSolTab}) is included into the entropy mode solution (the blue lines), since it does not propagate as all harmonics of the entropy wave. } Animations showing the evolution of separate modes and the development of full density perturbation (sum of entropy and slow MA modes) can be found in the supplementary materials.
	}
	\label{SolutFig}
\end{figure}

In the examples shown in Fig.~\ref{SolutFig}, we focus on the evolution of the entropy wave and one slow MA wave as the evolution of the other slow MA wave is symmetrical. Thus, for demonstration of the solution at $t=0$ we use half of the second term in Eq.~(\ref{rhon_delt_min_eq}).
For $t\ne 0$, the second slow MA wave escapes the $z$-domain shown in Fig.~\ref{SolutFig} and is therefore not visible.
We have varied the width $w$ of the initial signal (\ref{init_gaussian}) to obtain the most illustrative examples.

In the top row of Fig.~\ref{SolutFig}, we show evolution of the isothermal density perturbation (\ref{init_gaussian}) in region $I_{+}$ of the characteristic thermal misbalance times $\tau_{1,2}$, shown in Fig.~\ref{Ampdamp}. As it was demonstrated in Sec.~\ref{sec:parametric-analysis}, the entropy mode attenuates in this region of parameters. The slow MA mode, in turn, forms a propagating quasi-periodic pattern at some distance from the site of the initial perturbation {with the characteristic wavelength about 0.1$L$ and period about 0.07$L/c_{S0}$}, as a result of amplification of its harmonics and negative dispersion of the phase speed (see the top row of panels in Fig.~\ref{DispTab} and discussion in Sec.~\ref{sec:param-reg}).
{For example, for typical parameters of coronal loops with $L=200$\,Mm and $T_0 =1$--10\,MK, these wavelength and period are 20\,Mm and 40--120~s, respectively.}
The formation of a similar propagating slow MA wave train for a set of parameters from region $I_{+}$ was demonstrated in \citet{2019PhPl...26h2113Z} by numerical solution of the evolutionary Eq.~(\ref{lin_evol_eq}), i.e. without separating the total solution into the individual evolution of the entropy and slow MA waves.

The second row in Fig.~\ref{SolutFig} was obtained for a set of parameters $\tau_{1,2}$ corresponding to region $II_{+}$ in Fig.~\ref{Ampdamp} and the second row of panels in Fig.~\ref{DispTab}. In this case, both the entropy and slow MA waves decay, leading to the disappearance of the plasma density perturbation.  Moreover, the decaying and propagating slow MA pulse becomes strongly asymmetric. This effect is caused by a positive dispersion of the slow MA phase speed accompanied by stronger damping of higher harmonics. In addition, the slow MA pulse shape is affected by the effective excitation of the entropy mode in such a thermodynamically active plasma, that violates the symmetry in the partition of the initial perturbation energy across the spectrum. The decay of the propagating slow MA pulse in this regime was also shown numerically in \citet{2019PhPl...26h2113Z}, without discussion of the entropy mode behaviour.

The wave evolution in region $I_{-}$ of the parameters $\tau_{1,2}$ is shown by the third row of Fig.~\ref{SolutFig}. In this case, the entropy mode grows and leads to the formation of a localised plasma condensation at the site of the initial perturbation. The slow MA mode, in turn, runs away from the perturbation epicentre and decays. In the coronal context, for example, this regime could correspond to observations or numerical simulations of coronal rain formed in response to impulsive perturbations of the coronal mechanical and thermal equilibria, with poorly pronounced (or not pronounced at all) signatures of slow MA waves \citep[see e.g.][]{2020A&A...639A..20K}.

The opposite situation occurs in region $II_{-}$ (see the bottom row in Fig.~\ref{SolutFig} ), in which both entropy and slow MA modes are amplified. Thus, one should expect to observe an effective formation of localised plasma condensations by the growing entropy mode, accompanied by well developed and visible propagating slow MA waves. The exponential growth of both modes in this regime will breach the realm of the linear analysis when the perturbation amplitude becomes sufficiently large. However, taking additional dissipative processes, such as thermal conduction and viscosity, into account may suppress the wave growth rates or even stabilise the perturbations, thus extending the range of applicability of the developed linear theory. This issue will be addressed in the follow-up works.

\subsection{Partition of energy between entropy and slow MA modes} \label{Subsec 4_3}

In this section, we consider the question of the relative efficiency of the excitation of slow MA and entropy waves in a plasma with heating/cooling misbalance. The set of linear equations (\ref{const_mat_delt_min}) connecting the initial amplitude coefficients $C_{1n}$, $C_{2n}$, and $C_{3n}$ of the entropy and slow MA waves with the initial perturbation $ \rho_{in}(z,0)$, $\at{(\partial \rho(z,t) /  \partial t)}{t=0}$, and $ \at{(\partial^2 \rho(z,t) /  \partial t^2)}{t=0}$, can be treated as effective initial Fourier spectra of those waves, providing the distribution of the initial perturbation energy over the harmonic numbers $n$. Moreover, the partition of this initial energy between the entropy and slow MA waves and their harmonics is seen to directly depend on the properties of the plasma heat-loss function $Q(\rho, T )$ through the presence of the parameters $\tau_{1,2}$ on the LHS of Eq.~(\ref{const_mat_delt_min}). In other words, for some regimes of the misbalance (values of $\tau_{1,2}$) the entropy/slow MA waves could be excited with a higher/lower efficiency. We also note that the question of this energy partition makes sense only for the regimes of decaying waves, as otherwise the wave instability would cause the amplitude to grow exponentially independently of its initial value. Hence, in this section we consider region $II_+$ of the misbalance parameters $\tau_{1,2}$ (see Fig.~\ref{Ampdamp}), for which both modes decay and all slow MA harmonics propagate.


In this work, we do not discuss the distribution of the initial energy over the individual entropy and slow MA harmonics, but focus on the partition of the total (i.e. integrated over all harmonics) initial energy between the modes. Thus, the ratio of the total initial energies ${\cal E}_\mathrm{tot}$ and ${\cal A}_\mathrm{tot}$ gained by the entropy and slow MA modes, respectively, from the initial Gaussian pulse (\ref{init_gaussian}) can be estimated as
\begin{equation}\label{AmpRatio_Eq}
	\frac{{\cal E}_\mathrm{tot}}{{\cal A}_\mathrm{tot}} = \sum\limits_{n=1}^{\infty} C_{1n}^2/2 \sum\limits_{n=1}^{\infty}  C_{0n}^2,
\end{equation}
where
\begin{align}\label{ratio}
	\nonumber 
	&{\cal E}_\mathrm{tot} =	\sum\limits_{n=1}^{\infty} \left[\frac{I_{1n} \left(\omega_{AR}^2+\omega_{AI}^2\right) +I_{3n} }
	{\omega_{AR}^2+\left(\omega_{AI}+\omega_{EI}\right)^2}  \right]^2    ,\\
	&{\cal A}_\mathrm{tot} =	\sum\limits_{n=1}^{\infty} \frac{\omega_{EI}^2 I_{1n}^2 \left(\omega_{AR}^2+\omega_{AI}^2\right) - 2 \omega_{EI} \omega_{AI} I_{1n} I_{3n} + I_{3n}^2 }
	{\omega_{AR}^2\left[\omega_{AR}^2+\left(\omega_{AI}+\omega_{EI}\right)^2\right]}.  \nonumber
\end{align}
The values of the integrals $I_{1n}$ and $I_{3n}$ can be obtained either analytically or numerically using Eqs.~(\ref{init_integral_1})--(\ref{init_integral_3}) for the chosen initial conditions (\ref{eq:drho/dt})--(\ref{init_gaussian}).

\begin{figure}
	\begin{center}
		\includegraphics[width=0.99\linewidth]{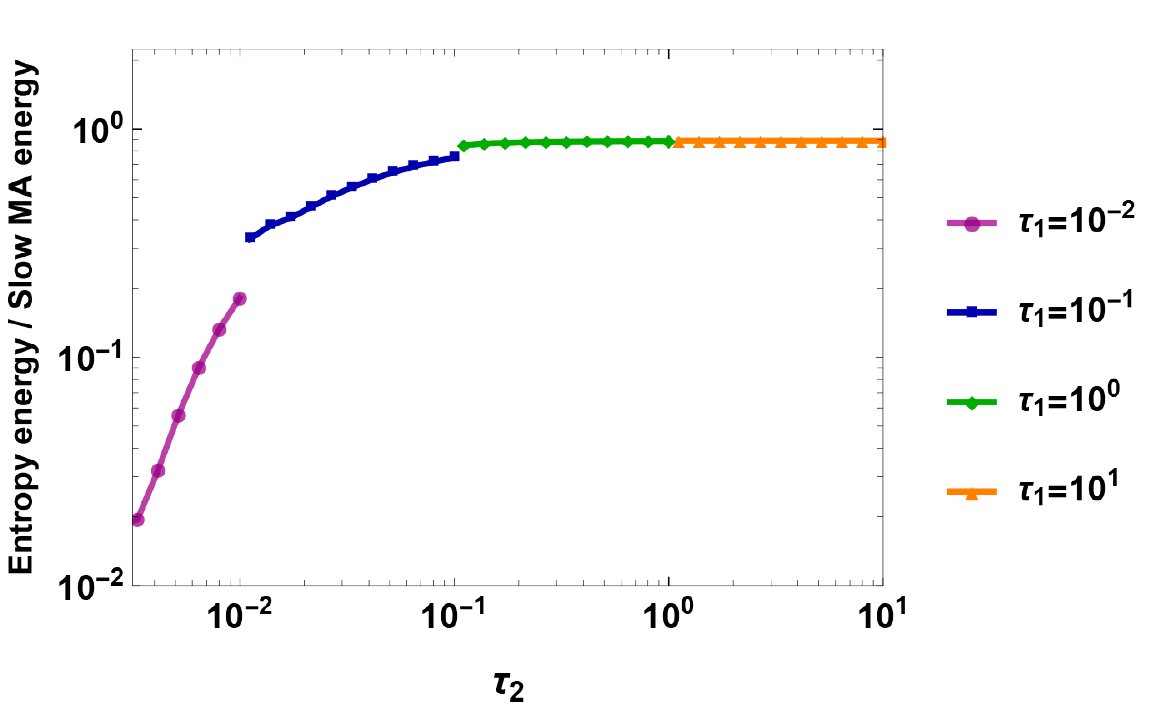}
	\end{center}
	\caption{Dependence of the entropy to slow MA total energy ratio ${{\cal E}_\mathrm{tot}}/{{\cal A}_\mathrm{tot}}$ (\ref{AmpRatio_Eq}) on the misbalance timescale $\tau_2$, calculated for 200 harmonics and for different values of the timescale $\tau_1$ in the parametric region $II_{+}$ (see Fig.~\ref{Ampdamp}), i.e. $\tau_{1}/9\leq\tau_{2}\leq\tau_{1}$. Both parameters $\tau_1$ and $\tau_2$ are normalised to the isothermal acoustic travel time along the loop, $L/c_{S0}$. 
	}
	\label{RatioFig}
\end{figure}

Figure~\ref{RatioFig} shows the dependence of the entropy to slow MA total energy ratio ${{\cal E}_\mathrm{tot}}/{{\cal A}_\mathrm{tot}}$ (\ref{AmpRatio_Eq}) on the characteristic misbalance times $\tau_{1,2}$. It is seen that for  $\tau_{1,2}\geq1$ in the considered region  ($II_+$), the total initial energy of slow MA waves is almost equal to the total initial energy of the entropy wave. For lower values of $\tau_{1,2}$, the ratio ${{\cal E}_\mathrm{tot}}/{{\cal A}_\mathrm{tot}}$ decreases, indicating a preferential excitation of slow MA waves in this regime of thermal misbalance.

\section{Summary and conclusions}
\label{sec:summary}

The obtained exact solution (Table~\ref{ExactSolTab}) provides a major advance in understanding the effects of thermal misbalance on the evolution of entropy and slow MA waves in the optically thin non-adiabatic plasma of the hot solar corona. It gives us comprehensive information about the spatio-temporal dynamics and dispersive properties of those waves, allows for the analysis of the energy partition between slow MA and entropy waves and across their harmonics. In particular, the developed theory allowed us to identify the regimes in which properties of slow MA and entropy waves get mixed through the mechanism of thermal misbalance. Below, we summarise the distinct misbalance-caused features of those waves, which have been revealed using the obtained exact solution. 



\begin{itemize}
	\item The exact analytical solution of the dispersion relation (\ref{cubic_eq}) of slow MA and entropy waves is obtained without assumption of weak amplification/attenuation \citep[cf. the previous works on thermal misbalance by][where the dispersion relation was solved under this assumptions]{2019PhPl...26h2113Z,2019A&A...628A.133K,2020arXiv201110437D}. {It allowed us to obtain the dependence of the phase speed and increment/decrement of entropy and slow MA waves on wavenumbers (see Fig.~\ref{DispTab}). In particular, it is shown that for some regimes of thermal misbalance the dependence of the slow MA wave phase speed on wavenumber can be even non-monotonic (see the third and fourth rows in Fig.~\ref{DispTab}).}
	\item The presented theory is developed for any values of the characterstic misbalance timescales $\tau_{1,2}$. Indeed, previous estimations of   $\tau_{1,2}$ for typical coronal conditions showed that they could be either positive or negative, depending on the specific form of the coronal heating function \citep{2020A&A...644A..33K}. Thus, the quantitative analysis performed in our work generalises and extends the qualitative picture of the impact of thermal misbalance on slow MA and entropy modes shown in Table~2 of \citet{2020A&A...644A..33K}.
	\item For specific regimes of thermal misbalance prescribed by the values of its characterstic timescales $\tau_{1,2}$, harmonics of slow MA mode may become non-propagating. The wavenumbers $k_{cr1}$ and $k_{cr2}$ (\ref{crit_eigen_eq1}) defining the non-propagating range have been obtained analytically. In this non-propagating regime, slow MA and entropy harmonics manifest mixed properties, so that the perturbation of a low-$\beta$ coronal plasma caused by these modes would look identically, as a non-propagating disturbance of the plasma density either growing or decaying with time.
	The longitudinal structuring of plasma density and temperature in initially uniform along the field coronal loops, caused by those non-propagating slow MA harmonics, may have different spatial scales depending on the values of the wavenumbers $k_{cr1}$ and $k_{cr2}$.
	Traditionally, the parallel non-uniformity of plasma structures in the solar atmosphere, affecting the dynamics of magnetoacoustic waves, is associated with the gravitational density stratification and/or divergence of the magnetic field lines with height \citep[see e.g.][]{2004A&A...415..705D, 2005ApJ...624L..57A, 2006A&A...460..893M, 2011ApJ...728...84B, 2012ApJ...748..110L, 2019A&A...625A.144R}.
	In this context, the non-propagation of essentially compressive slow MA harmonics offers an alternative mechanism for ceasing the uniformity of plasma along the loop, leading to the necessity to consider the interaction between the perturbing wave and the non-uniformity of the background plasma caused by this wave, and also indirectly affecting the waves elsewhere in the loop \citep[see e.g.][where the slow and fast kink magnetoacoustic oscillations were observed to co-exist in a bundle of coronal loops]{2017A&A...600A..37N}. A detailed analysis and validation of this effect and its consequences for coronal MHD seismology would require comprehensive numerical simulations, taking nonlinear and additional dissipative effects into account.
	\item {There are a number of different scenarios for the evolution of the initial perturbation caused by the combination of a frequency-dependent damping/amplification of slow MA and entropy modes and monotonic positive/negative or non-monotonic dispersion of slow MA waves, associated with the phenomenon of thermal misbalance. In particular, the entropy wave could decay and slow MA wave grows and develops into a propagating quasi-periodic slow MA wave train; both waves decay smoothing the initial perturbation out; entropy wave grows and forms a localised condensation of the plasma density, while slow MA wave runs away and decays; both waves grow. In the regime of formation of slow MA wave trains, their characteristic period for typical coronal conditions is shown to be {about a few minutes}, which coincides with typical periods of quasi-periodic pulsations often observed in the thermal emission from solar flares and usually interpreted in terms of the evolution of compressive MHD waves in the solar atmosphere \citep[see e.g.][]{2020STP.....6a...3K, 2018SSRv..214...45M, 2016SoPh..291.3143V}. The regime of formation of the localised plasma condensations by the instability of the entropy mode is of a clear importance in the context of coronal rain formation \citep[see e.g.][]{2010ApJ...716..154A, 2013ApJ...771L..29F}. In particular, the dynamics of a similar localised plasma condensation of an unspecified nature in a gravitationally stratified coronal loop was considered by \citet{2017A&A...602A..23K}, the initial formation of which could be self-consistently attributed to the instability of the entropy mode through the mechanism of thermal misbalance, developed in our work.
	The regime of misbalance causing both waves to decay could be used for the interpretation of rapid damping of standing, propagating, and sloshing slow MA waves in quiescent and active region coronal loops, polar plumes, and interplume regions \citep[][]{2009SSRv..149...65D, 2011SSRv..158..397W, 2011SSRv..158..267B, 2019ApJ...874L...1N}. The first successful attempts of applying the phenomenon of thermal misbalance to observations of rapidly decaying standing slow MA oscillations (also known as SUMER oscillations) in hot and dense loops in active regions were performed by \citet{2019A&A...628A.133K} and \citet{2021SoPh..296...20P}.}
	\item The initial distribution of energy in and between slow MA and entropy waves depends on the properties of the coronal heat-loss function $Q(\rho,T)$. It has been shown that for the initial perturbation of an isothermal nature the total initial energies are either equal or the most goes  into slow MA waves. In particular, for the characteristic misbalance timescales $\tau_{1,2}$ being about the acoustic travel time along the loop and higher, the ratio between the entropy and slow MA wave initial energies is about unity. For lower values of $\tau_{1,2}$, this ratio becomes  smaller. The estimation of this energy partition would also depend on the type of the initial perturbation. For example, excitation of slow MA waves in coronal loops by injected flows is often considered in modelling \citep[see e.g.][]{2012ApJ...754..111O, 2013ApJ...775L..23W, 2018AdSpR..61..645P}, and could lead to a different dependence of the entropy to slow MA total initial energy ratio on the misbalance parameters $\tau_{1,2}$. A more detailed analysis of this question constitutes another potentially important follow-up of this work.
\end{itemize}

Taking additional dissipative processes such as parallel thermal conduction and compressive viscosity into account in future works would broaden the applicability of the developed theory and obtained exact analytical solution. We expect that those additional dissipative processes of the coronal plasma may change the picture presented in this work quantitatively but not qualitatively, i.e. the aforesaid effects of thermal misbalance on the evolution of slow MA and entropy waves will be retained.
We also would like to mention that similarly to the process of parallel thermal conduction, the phenomenon of thermal misbalance could lead to the appearance of additional phase shifts between slow MA perturbations of the plasma density and temperature. The exact analytical solution obtained in this work is seen to be particularly useful for addressing this question in future works too.

\begin{acks}[Acknowledgements]                                                    
The work was supported in part  by the Ministry of Science and Higher Education of the Russian Federation by State assignment to educational and research institutions under Projects No. FSSS-2020-0014, 0023-2019-0003, and by Subsidy No.075-GZ/C3569/278.  D.Y.K. acknowledges support from the STFC consolidated grant ST/T000252/1.
\end{acks}

%


%
%

%
%
\bibliographystyle{spr-mp-sola}
\bibliography{sp_zav_2021}

\begin{thebibliography}{68}
\ifx\bisbn     \undefined \def\bisbn  #1{ISBN #1}\fi
\ifx\binits    \undefined \def\binits#1{#1}\fi
\ifx\bauthor   \undefined \def\bauthor#1{#1}\fi
\ifx\batitle   \undefined \def\batitle#1{#1}\fi
\ifx\bjtitle   \undefined \def\bjtitle#1{\textit{#1}}\fi
\ifx\bvolume   \undefined \def\bvolume#1{\textbf{#1}}\fi
\ifx\byear     \undefined \def\byear#1{#1}\fi
\ifx\bissue    \undefined \def\bissue#1{#1}\fi
\ifx\bfpage    \undefined \def\bfpage#1{#1}\fi
\ifx\blpage    \undefined \def\blpage #1{#1}\fi
\ifx\burl      \undefined \def\burl#1{\textsf{#1}}\fi
\ifx\href      \undefined \def\href#1#2{\textsf{#2}}\fi
\ifx\betal     \undefined \def\betal{\textit{et al.}}\fi
\ifx\bctitle   \undefined \def\bctitle#1{#1}\fi
\ifx\beditor   \undefined \def\beditor#1{#1}\fi
\ifx\bbtitle   \undefined \def\bbtitle#1{\textit{#1}}\fi
\ifx\bedition  \undefined \def\bedition#1{#1}\fi
\ifx\bseriesno \undefined \def\bseriesno#1{\textbf{#1}}\fi
\ifx\blocation \undefined \def\blocation#1{#1}\fi
\ifx\bsertitle \undefined \def\bsertitle#1{\textit{#1}}\fi
\ifx\bsnm      \undefined \def\bsnm#1{#1}\fi
\ifx\bsuffix   \undefined \def\bsuffix#1{#1}\fi
\ifx\bparticle \undefined \def\bparticle#1{#1}\fi
\ifx\barticle  \undefined \def\barticle#1{}\fi
\ifx\binstitute  \undefined \def\binstitute#1{#1}\fi
\ifx\bpublisher  \undefined \def\bpublisher#1{#1}\fi
\ifx\doiurl    \undefined \def\doiurl#1{\href{#1}{\textsf{DOI}}}\fi
\makeatletter
\def\safeHref#1#2#3{\in@{http}{#2}\ifin@\href{#2}{#3}\else\href{#1#2}{#3}\fi}
\makeatother
\ifx\adsurl    \undefined
  \def\adsurl#1{\safeHref{https://ui.adsabs.harvard.edu/abs/}{#1}{\textsf{ADS}}}\fi
\ifx\arxivurl  \undefined
  \def\arxivurl#1{\safeHref{http://arxiv.org/abs/}{#1}{\textsf{arXiv}}}\fi
\ifx\botherref \undefined \def\botherref#1{}\fi
\ifx\url       \undefined \def\url#1{\textsf{#1}}\fi
\ifx\bchapter  \undefined \def\bchapter#1{}\fi
\ifx\bbook     \undefined \def\bbook#1{}\fi
\ifx\bcomment  \undefined \def\bcomment#1{#1}\fi
\ifx\oauthor   \undefined \def\oauthor#1{#1}\fi
\ifx\citeauthoryear \undefined\def \citeauthoryear#1{#1}\fi
\def\endbibitem {}
\ifx\bconflocation  \undefined \def\bconflocation#1{#1} \fi

\bibitem[\protect\citeauthoryear{{Afanasyev} and
  {Nakariakov}}{2015}]{2015A&A...573A..32A}
\begin{barticle}
\bauthor{\bsnm{{Afanasyev}}, \binits{A.N.}},
\bauthor{\bsnm{{Nakariakov}}, \binits{V.M.}}:
\byear{2015},
\batitle{{Nonlinear slow magnetoacoustic waves in coronal plasma structures}}.
\bjtitle{\aap}
\bvolume{573},
\bfpage{A32}.
\doiurl{https://doi.org/10.1051/0004-6361/201424516}.
\adsurl{2015A&A...573A..32A}.
\end{barticle}
\endbibitem

\bibitem[\protect\citeauthoryear{{Andries}, {Arregui}, and
  {Goossens}}{2005}]{2005ApJ...624L..57A}
\begin{barticle}
\bauthor{\bsnm{{Andries}}, \binits{J.}},
\bauthor{\bsnm{{Arregui}}, \binits{I.}},
\bauthor{\bsnm{{Goossens}}, \binits{M.}}:
\byear{2005},
\batitle{{Determination of the Coronal Density Stratification from the
  Observation of Harmonic Coronal Loop Oscillations}}.
\bjtitle{\apjl}
\bvolume{624}(\bissue{1}),
\bfpage{L57}.
\doiurl{https://doi.org/10.1086/430347}.
\adsurl{2005ApJ...624L..57A}.
\end{barticle}
\endbibitem

\bibitem[\protect\citeauthoryear{{Antolin}}{2020}]{2020PPCF...62a4016A}
\begin{barticle}
\bauthor{\bsnm{{Antolin}}, \binits{P.}}:
\byear{2020},
\batitle{{Thermal instability and non-equilibrium in solar coronal loops: from
  coronal rain to long-period intensity pulsations}}.
\bjtitle{Plasma Physics and Controlled Fusion}
\bvolume{62}(\bissue{1}),
\bfpage{014016}.
\doiurl{https://doi.org/10.1088/1361-6587/ab5406}.
\adsurl{2020PPCF...62a4016A}.
\end{barticle}
\endbibitem

\bibitem[\protect\citeauthoryear{{Antolin}, {Shibata}, and
  {Vissers}}{2010}]{2010ApJ...716..154A}
\begin{barticle}
\bauthor{\bsnm{{Antolin}}, \binits{P.}},
\bauthor{\bsnm{{Shibata}}, \binits{K.}},
\bauthor{\bsnm{{Vissers}}, \binits{G.}}:
\byear{2010},
\batitle{{Coronal Rain as a Marker for Coronal Heating Mechanisms}}.
\bjtitle{\apj}
\bvolume{716}(\bissue{1}),
\bfpage{154}.
\doiurl{https://doi.org/10.1088/0004-637X/716/1/154}.
\adsurl{2010ApJ...716..154A}.
\end{barticle}
\endbibitem

\bibitem[\protect\citeauthoryear{{Banerjee} and {Krishna
  Prasad}}{2016}]{2016GMS...216..419B}
\begin{barticle}
\bauthor{\bsnm{{Banerjee}}, \binits{D.}},
\bauthor{\bsnm{{Krishna Prasad}}, \binits{S.}}:
\byear{2016},
\batitle{{MHD Waves in Coronal Holes}}.
\bjtitle{Washington DC American Geophysical Union Geophysical Monograph Series}
\bvolume{216},
\bfpage{419}.
\doiurl{https://doi.org/10.1002/9781119055006.ch24}.
\adsurl{2016GMS...216..419B}.
\end{barticle}
\endbibitem

\bibitem[\protect\citeauthoryear{{Banerjee}, {Gupta}, and
  {Teriaca}}{2011}]{2011SSRv..158..267B}
\begin{barticle}
\bauthor{\bsnm{{Banerjee}}, \binits{D.}},
\bauthor{\bsnm{{Gupta}}, \binits{G.R.}},
\bauthor{\bsnm{{Teriaca}}, \binits{L.}}:
\byear{2011},
\batitle{{Propagating MHD Waves in Coronal Holes}}.
\bjtitle{\ssr}
\bvolume{158}(\bissue{2-4}),
\bfpage{267}.
\doiurl{https://doi.org/10.1007/s11214-010-9698-z}.
\adsurl{2011SSRv..158..267B}.
\end{barticle}
\endbibitem

\bibitem[\protect\citeauthoryear{{Botha}
  \textit{et~al.}}{2011}]{2011ApJ...728...84B}
\begin{barticle}
\bauthor{\bsnm{{Botha}}, \binits{G.J.J.}},
\bauthor{\bsnm{{Arber}}, \binits{T.D.}},
\bauthor{\bsnm{{Nakariakov}}, \binits{V.M.}},
\bauthor{\bsnm{{Zhugzhda}}, \binits{Y.D.}}:
\byear{2011},
\batitle{{Chromospheric Resonances above Sunspot Umbrae}}.
\bjtitle{\apj}
\bvolume{728}(\bissue{2}),
\bfpage{84}.
\doiurl{https://doi.org/10.1088/0004-637X/728/2/84}.
\adsurl{2011ApJ...728...84B}.
\end{barticle}
\endbibitem

\bibitem[\protect\citeauthoryear{{Chin}
  \textit{et~al.}}{2010}]{2010PhPl...17c2107C}
\begin{barticle}
\bauthor{\bsnm{{Chin}}, \binits{R.}},
\bauthor{\bsnm{{Verwichte}}, \binits{E.}},
\bauthor{\bsnm{{Rowlands}}, \binits{G.}},
\bauthor{\bsnm{{Nakariakov}}, \binits{V.M.}}:
\byear{2010},
\batitle{{Self-organization of magnetoacoustic waves in a thermally unstable
  environment}}.
\bjtitle{Physics of Plasmas}
\bvolume{17}(\bissue{3}),
\bfpage{032107}.
\doiurl{https://doi.org/10.1063/1.3314721}.
\adsurl{2010PhPl...17c2107C}.
\end{barticle}
\endbibitem

\bibitem[\protect\citeauthoryear{{Claes} and
  {Keppens}}{2019}]{2019A&A...624A..96C}
\begin{barticle}
\bauthor{\bsnm{{Claes}}, \binits{N.}},
\bauthor{\bsnm{{Keppens}}, \binits{R.}}:
\byear{2019},
\batitle{{Thermal stability of magnetohydrodynamic modes in homogeneous
  plasmas}}.
\bjtitle{\aap}
\bvolume{624},
\bfpage{A96}.
\doiurl{https://doi.org/10.1051/0004-6361/201834699}.
\adsurl{2019A&A...624A..96C}.
\end{barticle}
\endbibitem

\bibitem[\protect\citeauthoryear{{De Moortel}}{2009}]{2009SSRv..149...65D}
\begin{barticle}
\bauthor{\bsnm{{De Moortel}}, \binits{I.}}:
\byear{2009},
\batitle{{Longitudinal Waves in Coronal Loops}}.
\bjtitle{\ssr}
\bvolume{149}(\bissue{1-4}),
\bfpage{65}.
\doiurl{https://doi.org/10.1007/s11214-009-9526-5}.
\adsurl{2009SSRv..149...65D}.
\end{barticle}
\endbibitem

\bibitem[\protect\citeauthoryear{{De Moortel} and
  {Hood}}{2003}]{2003A&A...408..755D}
\begin{barticle}
\bauthor{\bsnm{{De Moortel}}, \binits{I.}},
\bauthor{\bsnm{{Hood}}, \binits{A.W.}}:
\byear{2003},
\batitle{{The damping of slow MHD waves in solar coronal magnetic fields}}.
\bjtitle{\aap}
\bvolume{408},
\bfpage{755}.
\doiurl{https://doi.org/10.1051/0004-6361:20030984}.
\adsurl{2003A&A...408..755D}.
\end{barticle}
\endbibitem

\bibitem[\protect\citeauthoryear{{De Moortel} and
  {Hood}}{2004}]{2004A&A...415..705D}
\begin{barticle}
\bauthor{\bsnm{{De Moortel}}, \binits{I.}},
\bauthor{\bsnm{{Hood}}, \binits{A.W.}}:
\byear{2004},
\batitle{{The damping of slow MHD waves in solar coronal magnetic fields. II.
  The effect of gravitational stratification and field line divergence}}.
\bjtitle{\aap}
\bvolume{415},
\bfpage{705}.
\doiurl{https://doi.org/10.1051/0004-6361:20034233}.
\adsurl{2004A&A...415..705D}.
\end{barticle}
\endbibitem

\bibitem[\protect\citeauthoryear{{Duckenfield}, {Kolotkov}, and
  {Nakariakov}}{2021}]{2020arXiv201110437D}
\begin{barticle}
\bauthor{\bsnm{{Duckenfield}}, \binits{T.J.}},
\bauthor{\bsnm{{Kolotkov}}, \binits{D.Y.}},
\bauthor{\bsnm{{Nakariakov}}, \binits{V.M.}}:
\byear{2021},
\batitle{{The effect of the magnetic field on the damping of slow waves in the
  solar corona}}.
\bjtitle{\aap}
\bvolume{646},
\bfpage{A155}.
\doiurl{https://doi.org/10.1051/0004-6361/202039791}.
\adsurl{2021A&A...646A...}.
\end{barticle}
\endbibitem

\bibitem[\protect\citeauthoryear{{Fang}, {Xia}, and
  {Keppens}}{2013}]{2013ApJ...771L..29F}
\begin{barticle}
\bauthor{\bsnm{{Fang}}, \binits{X.}},
\bauthor{\bsnm{{Xia}}, \binits{C.}},
\bauthor{\bsnm{{Keppens}}, \binits{R.}}:
\byear{2013},
\batitle{{Multidimensional Modeling of Coronal Rain Dynamics}}.
\bjtitle{\apjl}
\bvolume{771}(\bissue{2}),
\bfpage{L29}.
\doiurl{https://doi.org/10.1088/2041-8205/771/2/L29}.
\adsurl{2013ApJ...771L..29F}.
\end{barticle}
\endbibitem

\bibitem[\protect\citeauthoryear{{Field}}{1965}]{1965ApJ...142..531F}
\begin{barticle}
\bauthor{\bsnm{{Field}}, \binits{G.B.}}:
\byear{1965},
\batitle{{Thermal Instability.}}
\bjtitle{\apj}
\bvolume{142},
\bfpage{531}.
\doiurl{https://doi.org/10.1086/148317}.
\adsurl{1965ApJ...142..531F}.
\end{barticle}
\endbibitem

\bibitem[\protect\citeauthoryear{{Goossens}, {Arregui}, and {Van
  Doorsselaere}}{2019}]{2019FrASS...6...20G}
\begin{barticle}
\bauthor{\bsnm{{Goossens}}, \binits{M.L.}},
\bauthor{\bsnm{{Arregui}}, \binits{I.}},
\bauthor{\bsnm{{Van Doorsselaere}}, \binits{T.}}:
\byear{2019},
\batitle{{Mixed properties of MHD waves in non-uniform plasmas}}.
\bjtitle{Frontiers in Astronomy and Space Sciences}
\bvolume{6},
\bfpage{20}.
\doiurl{https://doi.org/10.3389/fspas.2019.00020}.
\adsurl{2019FrASS...6...20G}.
\end{barticle}
\endbibitem

\bibitem[\protect\citeauthoryear{{Heyvaerts}}{1974}]{1974A&A....37...65H}
\begin{barticle}
\bauthor{\bsnm{{Heyvaerts}}, \binits{J.}}:
\byear{1974},
\batitle{{The thermal instability in a magnetohydrodynamic medium.}}
\bjtitle{\aap}
\bvolume{37}(\bissue{1}),
\bfpage{65}.
\adsurl{1974A&A....37...65H}.
\end{barticle}
\endbibitem

\bibitem[\protect\citeauthoryear{{Kaneko} and
  {Yokoyama}}{2017}]{2017ApJ...845...12K}
\begin{barticle}
\bauthor{\bsnm{{Kaneko}}, \binits{T.}},
\bauthor{\bsnm{{Yokoyama}}, \binits{T.}}:
\byear{2017},
\batitle{{Reconnection-Condensation Model for Solar Prominence Formation}}.
\bjtitle{\apj}
\bvolume{845}(\bissue{1}),
\bfpage{12}.
\doiurl{https://doi.org/10.3847/1538-4357/aa7d59}.
\adsurl{2017ApJ...845...12K}.
\end{barticle}
\endbibitem

\bibitem[\protect\citeauthoryear{{Kohutova} and
  {Verwichte}}{2017}]{2017A&A...602A..23K}
\begin{barticle}
\bauthor{\bsnm{{Kohutova}}, \binits{P.}},
\bauthor{\bsnm{{Verwichte}}, \binits{E.}}:
\byear{2017},
\batitle{{Dynamics of plasma condensations in a gravitationally stratified
  coronal loop}}.
\bjtitle{\aap}
\bvolume{602},
\bfpage{A23}.
\doiurl{https://doi.org/10.1051/0004-6361/201629912}.
\adsurl{2017A&A...602A..23K}.
\end{barticle}
\endbibitem

\bibitem[\protect\citeauthoryear{{Kohutova}
  \textit{et~al.}}{2020}]{2020A&A...639A..20K}
\begin{barticle}
\bauthor{\bsnm{{Kohutova}}, \binits{P.}},
\bauthor{\bsnm{{Antolin}}, \binits{P.}},
\bauthor{\bsnm{{Popovas}}, \binits{A.}},
\bauthor{\bsnm{{Szydlarski}}, \binits{M.}},
\bauthor{\bsnm{{Hansteen}}, \binits{V.H.}}:
\byear{2020},
\batitle{{Self-consistent 3D radiative magnetohydrodynamic simulations of
  coronal rain formation and evolution}}.
\bjtitle{\aap}
\bvolume{639},
\bfpage{A20}.
\doiurl{https://doi.org/10.1051/0004-6361/202037899}.
\adsurl{2020A&A...639A..20K}.
\end{barticle}
\endbibitem

\bibitem[\protect\citeauthoryear{{Kolotkov}, {Duckenfield}, and
  {Nakariakov}}{2020}]{2020A&A...644A..33K}
\begin{barticle}
\bauthor{\bsnm{{Kolotkov}}, \binits{D.Y.}},
\bauthor{\bsnm{{Duckenfield}}, \binits{T.J.}},
\bauthor{\bsnm{{Nakariakov}}, \binits{V.M.}}:
\byear{2020},
\batitle{{Seismological constraints on the solar coronal heating function}}.
\bjtitle{\aap}
\bvolume{644},
\bfpage{A33}.
\doiurl{https://doi.org/10.1051/0004-6361/202039095}.
\adsurl{2020A&A...644A..33K}.
\end{barticle}
\endbibitem

\bibitem[\protect\citeauthoryear{{Kolotkov}, {Nakariakov}, and
  {Zavershinskii}}{2019}]{2019A&A...628A.133K}
\begin{barticle}
\bauthor{\bsnm{{Kolotkov}}, \binits{D.Y.}},
\bauthor{\bsnm{{Nakariakov}}, \binits{V.M.}},
\bauthor{\bsnm{{Zavershinskii}}, \binits{D.I.}}:
\byear{2019},
\batitle{{Damping of slow magnetoacoustic oscillations by the misbalance
  between heating and cooling processes in the solar corona}}.
\bjtitle{\aap}
\bvolume{628},
\bfpage{A133}.
\doiurl{https://doi.org/10.1051/0004-6361/201936072}.
\adsurl{2019A&A...628A.133K}.
\end{barticle}
\endbibitem

\bibitem[\protect\citeauthoryear{{Krishna Prasad}, {Banerjee}, and {Van
  Doorsselaere}}{2014}]{2014ApJ...789..118K}
\begin{barticle}
\bauthor{\bsnm{{Krishna Prasad}}, \binits{S.}},
\bauthor{\bsnm{{Banerjee}}, \binits{D.}},
\bauthor{\bsnm{{Van Doorsselaere}}, \binits{T.}}:
\byear{2014},
\batitle{{Frequency-dependent Damping in Propagating Slow Magneto-acoustic
  Waves}}.
\bjtitle{\apj}
\bvolume{789}(\bissue{2}),
\bfpage{118}.
\doiurl{https://doi.org/10.1088/0004-637X/789/2/118}.
\adsurl{2014ApJ...789..118K}.
\end{barticle}
\endbibitem

\bibitem[\protect\citeauthoryear{{Krishna Prasad}, {Jess}, and {Van
  Doorsselaere}}{2019}]{2019FrASS...6...57S}
\begin{barticle}
\bauthor{\bsnm{{Krishna Prasad}}, \binits{S.}},
\bauthor{\bsnm{{Jess}}, \binits{D.B.}},
\bauthor{\bsnm{{Van Doorsselaere}}, \binits{T.}}:
\byear{2019},
\batitle{{The temperature-dependent damping of propagating slow magnetoacoustic
  waves}}.
\bjtitle{Frontiers in Astronomy and Space Sciences}
\bvolume{6},
\bfpage{57}.
\doiurl{https://doi.org/10.3389/fspas.2019.00057}.
\adsurl{2019FrASS...6...57S}.
\end{barticle}
\endbibitem

\bibitem[\protect\citeauthoryear{{Krishna Prasad}
  \textit{et~al.}}{2018}]{2018ApJ...868..149K}
\begin{barticle}
\bauthor{\bsnm{{Krishna Prasad}}, \binits{S.}},
\bauthor{\bsnm{{Raes}}, \binits{J.O.}},
\bauthor{\bsnm{{Van Doorsselaere}}, \binits{T.}},
\bauthor{\bsnm{{Magyar}}, \binits{N.}},
\bauthor{\bsnm{{Jess}}, \binits{D.B.}}:
\byear{2018},
\batitle{{The Polytropic Index of Solar Coronal Plasma in Sunspot Fan Loops and
  Its Temperature Dependence}}.
\bjtitle{\apj}
\bvolume{868}(\bissue{2}),
\bfpage{149}.
\doiurl{https://doi.org/10.3847/1538-4357/aae9f5}.
\adsurl{2018ApJ...868..149K}.
\end{barticle}
\endbibitem

\bibitem[\protect\citeauthoryear{{Kupriyanova}
  \textit{et~al.}}{2020}]{2020STP.....6a...3K}
\begin{barticle}
\bauthor{\bsnm{{Kupriyanova}}, \binits{E.}},
\bauthor{\bsnm{{Kolotkov}}, \binits{D.}},
\bauthor{\bsnm{{Nakariakov}}, \binits{V.}},
\bauthor{\bsnm{{Kaufman}}, \binits{A.}}:
\byear{2020},
\batitle{{Quasi-Periodic Pulsations in Solar and Stellar Flares. Review}}.
\bjtitle{Solar-Terrestrial Physics}
\bvolume{6}(\bissue{1}),
\bfpage{3}.
\doiurl{https://doi.org/10.12737/stp-61202001}.
\adsurl{2020STP.....6a...3K}.
\end{barticle}
\endbibitem

\bibitem[\protect\citeauthoryear{{Luna-Cardozo}, {Verth}, and
  {Erd{\'e}lyi}}{2012}]{2012ApJ...748..110L}
\begin{barticle}
\bauthor{\bsnm{{Luna-Cardozo}}, \binits{M.}},
\bauthor{\bsnm{{Verth}}, \binits{G.}},
\bauthor{\bsnm{{Erd{\'e}lyi}}, \binits{R.}}:
\byear{2012},
\batitle{{Longitudinal Oscillations in Density Stratified and Expanding Solar
  Waveguides}}.
\bjtitle{\apj}
\bvolume{748}(\bissue{2}),
\bfpage{110}.
\doiurl{https://doi.org/10.1088/0004-637X/748/2/110}.
\adsurl{2012ApJ...748..110L}.
\end{barticle}
\endbibitem

\bibitem[\protect\citeauthoryear{{Mandal}
  \textit{et~al.}}{2016}]{2016ApJ...820...13M}
\begin{barticle}
\bauthor{\bsnm{{Mandal}}, \binits{S.}},
\bauthor{\bsnm{{Magyar}}, \binits{N.}},
\bauthor{\bsnm{{Yuan}}, \binits{D.}},
\bauthor{\bsnm{{Van Doorsselaere}}, \binits{T.}},
\bauthor{\bsnm{{Banerjee}}, \binits{D.}}:
\byear{2016},
\batitle{{Forward Modeling of Propagating Slow Waves in Coronal Loops and Their
  Frequency-dependent Damping}}.
\bjtitle{\apj}
\bvolume{820}(\bissue{1}),
\bfpage{13}.
\doiurl{https://doi.org/10.3847/0004-637X/820/1/13}.
\adsurl{2016ApJ...820...13M}.
\end{barticle}
\endbibitem

\bibitem[\protect\citeauthoryear{{McEwan}
  \textit{et~al.}}{2006}]{2006A&A...460..893M}
\begin{barticle}
\bauthor{\bsnm{{McEwan}}, \binits{M.P.}},
\bauthor{\bsnm{{Donnelly}}, \binits{G.R.}},
\bauthor{\bsnm{{D{\'\i}az}}, \binits{A.J.}},
\bauthor{\bsnm{{Roberts}}, \binits{B.}}:
\byear{2006},
\batitle{{On the period ratio P$_{1}$/2P$_{2}$ in the oscillations of coronal
  loops}}.
\bjtitle{\aap}
\bvolume{460}(\bissue{3}),
\bfpage{893}.
\doiurl{https://doi.org/10.1051/0004-6361:20065313}.
\adsurl{2006A&A...460..893M}.
\end{barticle}
\endbibitem

\bibitem[\protect\citeauthoryear{{McLaughlin}
  \textit{et~al.}}{2018}]{2018SSRv..214...45M}
\begin{barticle}
\bauthor{\bsnm{{McLaughlin}}, \binits{J.A.}},
\bauthor{\bsnm{{Nakariakov}}, \binits{V.M.}},
\bauthor{\bsnm{{Dominique}}, \binits{M.}},
\bauthor{\bsnm{{Jel{\'\i}nek}}, \binits{P.}},
\bauthor{\bsnm{{Takasao}}, \binits{S.}}:
\byear{2018},
\batitle{{Modelling Quasi-Periodic Pulsations in Solar and Stellar Flares}}.
\bjtitle{\ssr}
\bvolume{214}(\bissue{1}),
\bfpage{45}.
\doiurl{https://doi.org/10.1007/s11214-018-0478-5}.
\adsurl{2018SSRv..214...45M}.
\end{barticle}
\endbibitem

\bibitem[\protect\citeauthoryear{Molevich, Zavershinskiy, and
  Ryashchikov}{2016}]{Molevich2016191}
\begin{barticle}
\bauthor{\bsnm{Molevich}, \binits{N.E.}},
\bauthor{\bsnm{Zavershinskiy}, \binits{D.I.}},
\bauthor{\bsnm{Ryashchikov}, \binits{D.S.}}:
\byear{2016},
\batitle{Investigation of the mhd wave dynamics in thermally unstable plasma}.
\bjtitle{Magnetohydrodynamics}
\bvolume{52}(\bissue{1}),
\bfpage{191}.
\bcomment{cited By 11}.
\doiurl{https://doi.org/10.22364/mhd.52.1.22}.
\end{barticle}
\endbibitem

\bibitem[\protect\citeauthoryear{Molevich and Oraevskii}{1988}]{Molevich88}
\begin{barticle}
\bauthor{\bsnm{Molevich}, \binits{N.E.}},
\bauthor{\bsnm{Oraevskii}, \binits{A.N.}}:
\byear{1988},
\batitle{Second viscosity in thermodynamically nonequilibrium media}.
\bjtitle{Zh. Eksp. Teor. Fiz}
\bvolume{94},
\bfpage{128}.
\bcomment{[J. Exp. Theor. Phys. {\bf 67}, 504 (1988)]}.
\end{barticle}
\endbibitem

\bibitem[\protect\citeauthoryear{{Molevich} and
  {Ryashchikov}}{2020}]{2020TePhL..46..637M}
\begin{barticle}
\bauthor{\bsnm{{Molevich}}, \binits{N.E.}},
\bauthor{\bsnm{{Ryashchikov}}, \binits{D.S.}}:
\byear{2020},
\batitle{{Autowave Pulse in a Medium with the Heating/Cooling Misbalance and an
  Arbitrary Thermal Dispersion}}.
\bjtitle{Technical Physics Letters}
\bvolume{46}(\bissue{7}),
\bfpage{637}.
\doiurl{https://doi.org/10.1134/S1063785020070123}.
\adsurl{2020TePhL..46..637M}.
\end{barticle}
\endbibitem

\bibitem[\protect\citeauthoryear{{Molevich}
  \textit{et~al.}}{2011}]{2011Ap&SS.334...35M}
\begin{barticle}
\bauthor{\bsnm{{Molevich}}, \binits{N.E.}},
\bauthor{\bsnm{{Zavershinsky}}, \binits{D.I.}},
\bauthor{\bsnm{{Galimov}}, \binits{R.N.}},
\bauthor{\bsnm{{Makaryan}}, \binits{V.G.}}:
\byear{2011},
\batitle{{Traveling self-sustained structures in interstellar clouds with the
  isentropic instability}}.
\bjtitle{\apss}
\bvolume{334}(\bissue{1}),
\bfpage{35}.
\doiurl{https://doi.org/10.1007/s10509-011-0683-0}.
\adsurl{2011Ap&SS.334...35M}.
\end{barticle}
\endbibitem

\bibitem[\protect\citeauthoryear{{Murawski}, {Zaqarashvili}, and
  {Nakariakov}}{2011}]{2011A&A...533A..18M}
\begin{barticle}
\bauthor{\bsnm{{Murawski}}, \binits{K.}},
\bauthor{\bsnm{{Zaqarashvili}}, \binits{T.V.}},
\bauthor{\bsnm{{Nakariakov}}, \binits{V.M.}}:
\byear{2011},
\batitle{{Entropy mode at a magnetic null point as a possible tool for indirect
  observation of nanoflares in the solar corona}}.
\bjtitle{\aap}
\bvolume{533},
\bfpage{A18}.
\doiurl{https://doi.org/10.1051/0004-6361/201116942}.
\adsurl{2011A&A...533A..18M}.
\end{barticle}
\endbibitem

\bibitem[\protect\citeauthoryear{{Nakariakov} and
  {Kolotkov}}{2020}]{2020ARA&A..58..441N}
\begin{barticle}
\bauthor{\bsnm{{Nakariakov}}, \binits{V.M.}},
\bauthor{\bsnm{{Kolotkov}}, \binits{D.Y.}}:
\byear{2020},
\batitle{{Magnetohydrodynamic Waves in the Solar Corona}}.
\bjtitle{\araa}
\bvolume{58},
\bfpage{441}.
\doiurl{https://doi.org/10.1146/annurev-astro-032320-042940}.
\adsurl{2020ARA&A..58..441N}.
\end{barticle}
\endbibitem

\bibitem[\protect\citeauthoryear{{Nakariakov}, {Mendoza-Brice{\~n}o}, and
  {Ib{\'a}{\~n}ez S.}}{2000}]{2000ApJ...528..767N}
\begin{barticle}
\bauthor{\bsnm{{Nakariakov}}, \binits{V.M.}},
\bauthor{\bsnm{{Mendoza-Brice{\~n}o}}, \binits{C.A.}},
\bauthor{\bsnm{{Ib{\'a}{\~n}ez S.}}, \binits{M.H.}}:
\byear{2000},
\batitle{{Magnetoacoustic Waves of Small Amplitude in Optically Thin
  Quasi-isentropic Plasmas}}.
\bjtitle{\apj}
\bvolume{528}(\bissue{2}),
\bfpage{767}.
\doiurl{https://doi.org/10.1086/308195}.
\adsurl{2000ApJ...528..767N}.
\end{barticle}
\endbibitem

\bibitem[\protect\citeauthoryear{{Nakariakov}
  \textit{et~al.}}{2017}]{2017ApJ...849...62N}
\begin{barticle}
\bauthor{\bsnm{{Nakariakov}}, \binits{V.M.}},
\bauthor{\bsnm{{Afanasyev}}, \binits{A.N.}},
\bauthor{\bsnm{{Kumar}}, \binits{S.}},
\bauthor{\bsnm{{Moon}}, \binits{Y.-J.}}:
\byear{2017},
\batitle{{Effect of Local Thermal Equilibrium Misbalance on Long-wavelength
  Slow Magnetoacoustic Waves}}.
\bjtitle{\apj}
\bvolume{849}(\bissue{1}),
\bfpage{62}.
\doiurl{https://doi.org/10.3847/1538-4357/aa8ea3}.
\adsurl{2017ApJ...849...62N}.
\end{barticle}
\endbibitem

\bibitem[\protect\citeauthoryear{{Nakariakov}
  \textit{et~al.}}{2019}]{2019ApJ...874L...1N}
\begin{barticle}
\bauthor{\bsnm{{Nakariakov}}, \binits{V.M.}},
\bauthor{\bsnm{{Kosak}}, \binits{M.K.}},
\bauthor{\bsnm{{Kolotkov}}, \binits{D.Y.}},
\bauthor{\bsnm{{Anfinogentov}}, \binits{S.A.}},
\bauthor{\bsnm{{Kumar}}, \binits{P.}},
\bauthor{\bsnm{{Moon}}, \binits{Y.-J.}}:
\byear{2019},
\batitle{{Properties of Slow Magnetoacoustic Oscillations of Solar Coronal
  Loops by Multi-instrumental Observations}}.
\bjtitle{\apjl}
\bvolume{874}(\bissue{1}),
\bfpage{L1}.
\doiurl{https://doi.org/10.3847/2041-8213/ab0c9f}.
\adsurl{2019ApJ...874L...1N}.
\end{barticle}
\endbibitem

\bibitem[\protect\citeauthoryear{{Nistic{\`o}}
  \textit{et~al.}}{2017}]{2017A&A...600A..37N}
\begin{barticle}
\bauthor{\bsnm{{Nistic{\`o}}}, \binits{G.}},
\bauthor{\bsnm{{Polito}}, \binits{V.}},
\bauthor{\bsnm{{Nakariakov}}, \binits{V.M.}},
\bauthor{\bsnm{{Del Zanna}}, \binits{G.}}:
\byear{2017},
\batitle{{Multi-instrument observations of a failed flare eruption associated
  with MHD waves in a loop bundle}}.
\bjtitle{\aap}
\bvolume{600},
\bfpage{A37}.
\doiurl{https://doi.org/10.1051/0004-6361/201629324}.
\adsurl{2017A&A...600A..37N}.
\end{barticle}
\endbibitem

\bibitem[\protect\citeauthoryear{{Ofman} and
  {Wang}}{2002}]{2002ApJ...580L..85O}
\begin{barticle}
\bauthor{\bsnm{{Ofman}}, \binits{L.}},
\bauthor{\bsnm{{Wang}}, \binits{T.}}:
\byear{2002},
\batitle{{Hot Coronal Loop Oscillations Observed by SUMER: Slow Magnetosonic
  Wave Damping by Thermal Conduction}}.
\bjtitle{\apjl}
\bvolume{580}(\bissue{1}),
\bfpage{L85}.
\doiurl{https://doi.org/10.1086/345548}.
\adsurl{2002ApJ...580L..85O}.
\end{barticle}
\endbibitem

\bibitem[\protect\citeauthoryear{{Ofman}, {Wang}, and
  {Davila}}{2012}]{2012ApJ...754..111O}
\begin{barticle}
\bauthor{\bsnm{{Ofman}}, \binits{L.}},
\bauthor{\bsnm{{Wang}}, \binits{T.J.}},
\bauthor{\bsnm{{Davila}}, \binits{J.M.}}:
\byear{2012},
\batitle{{Slow Magnetosonic Waves and Fast Flows in Active Region Loops}}.
\bjtitle{\apj}
\bvolume{754}(\bissue{2}),
\bfpage{111}.
\doiurl{https://doi.org/10.1088/0004-637X/754/2/111}.
\adsurl{2012ApJ...754..111O}.
\end{barticle}
\endbibitem

\bibitem[\protect\citeauthoryear{{Owen}, {De Moortel}, and
  {Hood}}{2009}]{2009A&A...494..339O}
\begin{barticle}
\bauthor{\bsnm{{Owen}}, \binits{N.R.}},
\bauthor{\bsnm{{De Moortel}}, \binits{I.}},
\bauthor{\bsnm{{Hood}}, \binits{A.W.}}:
\byear{2009},
\batitle{{Forward modelling to determine the observational signatures of
  propagating slow waves for TRACE, SoHO/CDS, and Hinode/EIS}}.
\bjtitle{\aap}
\bvolume{494}(\bissue{1}),
\bfpage{339}.
\doiurl{https://doi.org/10.1051/0004-6361:200810828}.
\adsurl{2009A&A...494..339O}.
\end{barticle}
\endbibitem

\bibitem[\protect\citeauthoryear{{Polyanin} and
  {Zaitsev}}{1995}]{1995heso.book.....P}
\begin{bbook}
\bauthor{\bsnm{{Polyanin}}, \binits{A.D.}},
\bauthor{\bsnm{{Zaitsev}}, \binits{V.F.}}:
\byear{1995},
\bbtitle{{Handbook of exact solutions for ordinary differential equations}}.
\adsurl{1995heso.book.....P}.
\end{bbook}
\endbibitem

\bibitem[\protect\citeauthoryear{{Prasad}, {Srivastava}, and
  {Wang}}{2021}]{2021SoPh..296...20P}
\begin{barticle}
\bauthor{\bsnm{{Prasad}}, \binits{A.}},
\bauthor{\bsnm{{Srivastava}}, \binits{A.K.}},
\bauthor{\bsnm{{Wang}}, \binits{T.J.}}:
\byear{2021},
\batitle{{Role of Compressive Viscosity and Thermal Conductivity on the Damping
  of Slow Waves in Coronal Loops with and Without Heating-Cooling Imbalance}}.
\bjtitle{\solphys}
\bvolume{296}(\bissue{1}),
\bfpage{20}.
\doiurl{https://doi.org/10.1007/s11207-021-01764-x}.
\adsurl{2021SoPh..296...20P}.
\end{barticle}
\endbibitem

\bibitem[\protect\citeauthoryear{{Provornikova}, {Ofman}, and
  {Wang}}{2018}]{2018AdSpR..61..645P}
\begin{barticle}
\bauthor{\bsnm{{Provornikova}}, \binits{E.}},
\bauthor{\bsnm{{Ofman}}, \binits{L.}},
\bauthor{\bsnm{{Wang}}, \binits{T.}}:
\byear{2018},
\batitle{{Excitation of flare-induced waves in coronal loops and the effects of
  radiative cooling}}.
\bjtitle{Advances in Space Research}
\bvolume{61}(\bissue{2}),
\bfpage{645}.
\doiurl{https://doi.org/10.1016/j.asr.2017.07.042}.
\adsurl{2018AdSpR..61..645P}.
\end{barticle}
\endbibitem

\bibitem[\protect\citeauthoryear{{Reale}}{2016}]{2016ApJ...826L..20R}
\begin{barticle}
\bauthor{\bsnm{{Reale}}, \binits{F.}}:
\byear{2016},
\batitle{{Plasma Sloshing in Pulse-heated Solar and Stellar Coronal Loops}}.
\bjtitle{\apjl}
\bvolume{826}(\bissue{2}),
\bfpage{L20}.
\doiurl{https://doi.org/10.3847/2041-8205/826/2/L20}.
\adsurl{2016ApJ...826L..20R}.
\end{barticle}
\endbibitem

\bibitem[\protect\citeauthoryear{{Reale}
  \textit{et~al.}}{2018}]{2018ApJ...856...51R}
\begin{barticle}
\bauthor{\bsnm{{Reale}}, \binits{F.}},
\bauthor{\bsnm{{Lopez-Santiago}}, \binits{J.}},
\bauthor{\bsnm{{Flaccomio}}, \binits{E.}},
\bauthor{\bsnm{{Petralia}}, \binits{A.}},
\bauthor{\bsnm{{Sciortino}}, \binits{S.}}:
\byear{2018},
\batitle{{X-Ray Flare Oscillations Track Plasma Sloshing along Star-disk
  Magnetic Tubes in the Orion Star-forming Region}}.
\bjtitle{\apj}
\bvolume{856}(\bissue{1}),
\bfpage{51}.
\doiurl{https://doi.org/10.3847/1538-4357/aaaf1f}.
\adsurl{2018ApJ...856...51R}.
\end{barticle}
\endbibitem

\bibitem[\protect\citeauthoryear{{Reale}
  \textit{et~al.}}{2019}]{2019ApJ...884..131R}
\begin{barticle}
\bauthor{\bsnm{{Reale}}, \binits{F.}},
\bauthor{\bsnm{{Testa}}, \binits{P.}},
\bauthor{\bsnm{{Petralia}}, \binits{A.}},
\bauthor{\bsnm{{Kolotkov}}, \binits{D.Y.}}:
\byear{2019},
\batitle{{Large-amplitude Quasiperiodic Pulsations as Evidence of Impulsive
  Heating in Hot Transient Loop Systems Detected in the EUV with SDO/AIA}}.
\bjtitle{\apj}
\bvolume{884}(\bissue{2}),
\bfpage{131}.
\doiurl{https://doi.org/10.3847/1538-4357/ab4270}.
\adsurl{2019ApJ...884..131R}.
\end{barticle}
\endbibitem

\bibitem[\protect\citeauthoryear{{Riedl}, {Van Doorsselaere}, and
  {Santamaria}}{2019}]{2019A&A...625A.144R}
\begin{barticle}
\bauthor{\bsnm{{Riedl}}, \binits{J.M.}},
\bauthor{\bsnm{{Van Doorsselaere}}, \binits{T.}},
\bauthor{\bsnm{{Santamaria}}, \binits{I.C.}}:
\byear{2019},
\batitle{{Wave modes excited by photospheric p-modes and mode conversion in a
  multi-loop system}}.
\bjtitle{\aap}
\bvolume{625},
\bfpage{A144}.
\doiurl{https://doi.org/10.1051/0004-6361/201935393}.
\adsurl{2019A&A...625A.144R}.
\end{barticle}
\endbibitem

\bibitem[\protect\citeauthoryear{{Ruderman}}{2006}]{2006RSPTA.364..485R}
\begin{barticle}
\bauthor{\bsnm{{Ruderman}}, \binits{M.S.}}:
\byear{2006},
\batitle{{Nonlinear waves in the solar atmosphere}}.
\bjtitle{Philosophical Transactions of the Royal Society of London Series A}
\bvolume{364}(\bissue{1839}),
\bfpage{485}.
\doiurl{https://doi.org/10.1098/rsta.2005.1712}.
\adsurl{2006RSPTA.364..485R}.
\end{barticle}
\endbibitem

\bibitem[\protect\citeauthoryear{Ryashchikov, Molevich, and
  Zavershinskii}{2017}]{Ryashchikov2017416}
\begin{barticle}
\bauthor{\bsnm{Ryashchikov}, \binits{D.S.}},
\bauthor{\bsnm{Molevich}, \binits{N.E.}},
\bauthor{\bsnm{Zavershinskii}, \binits{D.I.}}:
\byear{2017},
\batitle{Characteristic times of acoustic and condensation instability in
  heat-releasing gas media}.
\bjtitle{Procedia Engineering}
\bvolume{176},
\bfpage{416}.
\bcomment{cited By 1}.
\doiurl{https://doi.org/10.1016/j.proeng.2017.02.340}.
\end{barticle}
\endbibitem

\bibitem[\protect\citeauthoryear{{Selwa}, {Murawski}, and
  {Solanki}}{2005}]{2005A&A...436..701S}
\begin{barticle}
\bauthor{\bsnm{{Selwa}}, \binits{M.}},
\bauthor{\bsnm{{Murawski}}, \binits{K.}},
\bauthor{\bsnm{{Solanki}}, \binits{S.K.}}:
\byear{2005},
\batitle{{Excitation and damping of slow magnetosonic standing waves in a solar
  coronal loop}}.
\bjtitle{\aap}
\bvolume{436}(\bissue{2}),
\bfpage{701}.
\doiurl{https://doi.org/10.1051/0004-6361:20042319}.
\adsurl{2005A&A...436..701S}.
\end{barticle}
\endbibitem

\bibitem[\protect\citeauthoryear{{Somov}, {Dzhalilov}, and
  {Staude}}{2007}]{2007AstL...33..309S}
\begin{barticle}
\bauthor{\bsnm{{Somov}}, \binits{B.V.}},
\bauthor{\bsnm{{Dzhalilov}}, \binits{N.S.}},
\bauthor{\bsnm{{Staude}}, \binits{J.}}:
\byear{2007},
\batitle{{Peculiarities of entropy and magnetosonic waves in optically thin
  cosmic plasma}}.
\bjtitle{Astronomy Letters}
\bvolume{33}(\bissue{5}),
\bfpage{309}.
\doiurl{https://doi.org/10.1134/S1063773707050040}.
\adsurl{2007AstL...33..309S}.
\end{barticle}
\endbibitem

\bibitem[\protect\citeauthoryear{{Van Doorsselaere}, {Kupriyanova}, and
  {Yuan}}{2016}]{2016SoPh..291.3143V}
\begin{barticle}
\bauthor{\bsnm{{Van Doorsselaere}}, \binits{T.}},
\bauthor{\bsnm{{Kupriyanova}}, \binits{E.G.}},
\bauthor{\bsnm{{Yuan}}, \binits{D.}}:
\byear{2016},
\batitle{{Quasi-periodic Pulsations in Solar and Stellar Flares: An Overview of
  Recent Results (Invited Review)}}.
\bjtitle{\solphys}
\bvolume{291}(\bissue{11}),
\bfpage{3143}.
\doiurl{https://doi.org/10.1007/s11207-016-0977-z}.
\adsurl{2016SoPh..291.3143V}.
\end{barticle}
\endbibitem

\bibitem[\protect\citeauthoryear{{Van Doorsselaere}
  \textit{et~al.}}{2011}]{2011ApJ...727L..32V}
\begin{barticle}
\bauthor{\bsnm{{Van Doorsselaere}}, \binits{T.}},
\bauthor{\bsnm{{Wardle}}, \binits{N.}},
\bauthor{\bsnm{{Del Zanna}}, \binits{G.}},
\bauthor{\bsnm{{Jansari}}, \binits{K.}},
\bauthor{\bsnm{{Verwichte}}, \binits{E.}},
\bauthor{\bsnm{{Nakariakov}}, \binits{V.M.}}:
\byear{2011},
\batitle{{The First Measurement of the Adiabatic Index in the Solar Corona
  Using Time-dependent Spectroscopy of Hinode/EIS Observations}}.
\bjtitle{\apjl}
\bvolume{727}(\bissue{2}),
\bfpage{L32}.
\doiurl{https://doi.org/10.1088/2041-8205/727/2/L32}.
\adsurl{2011ApJ...727L..32V}.
\end{barticle}
\endbibitem

\bibitem[\protect\citeauthoryear{{Verwichte}
  \textit{et~al.}}{2008}]{2008ApJ...685.1286V}
\begin{barticle}
\bauthor{\bsnm{{Verwichte}}, \binits{E.}},
\bauthor{\bsnm{{Haynes}}, \binits{M.}},
\bauthor{\bsnm{{Arber}}, \binits{T.D.}},
\bauthor{\bsnm{{Brady}}, \binits{C.S.}}:
\byear{2008},
\batitle{{Damping of Slow MHD Coronal Loop Oscillations by Shocks}}.
\bjtitle{\apj}
\bvolume{685}(\bissue{2}),
\bfpage{1286}.
\doiurl{https://doi.org/10.1086/591077}.
\adsurl{2008ApJ...685.1286V}.
\end{barticle}
\endbibitem

\bibitem[\protect\citeauthoryear{{Wang}}{2011}]{2011SSRv..158..397W}
\begin{barticle}
\bauthor{\bsnm{{Wang}}, \binits{T.}}:
\byear{2011},
\batitle{{Standing Slow-Mode Waves in Hot Coronal Loops: Observations,
  Modeling, and Coronal Seismology}}.
\bjtitle{\ssr}
\bvolume{158}(\bissue{2-4}),
\bfpage{397}.
\doiurl{https://doi.org/10.1007/s11214-010-9716-1}.
\adsurl{2011SSRv..158..397W}.
\end{barticle}
\endbibitem

\bibitem[\protect\citeauthoryear{{Wang} and
  {Ofman}}{2019}]{2019ApJ...886....2W}
\begin{barticle}
\bauthor{\bsnm{{Wang}}, \binits{T.}},
\bauthor{\bsnm{{Ofman}}, \binits{L.}}:
\byear{2019},
\batitle{{Determination of Transport Coefficients by Coronal Seismology of
  Flare-induced Slow-mode Waves: Numerical Parametric Study of a 1D Loop
  Model}}.
\bjtitle{\apj}
\bvolume{886}(\bissue{1}),
\bfpage{2}.
\doiurl{https://doi.org/10.3847/1538-4357/ab478f}.
\adsurl{2019ApJ...886....2W}.
\end{barticle}
\endbibitem

\bibitem[\protect\citeauthoryear{{Wang}, {Ofman}, and
  {Davila}}{2013}]{2013ApJ...775L..23W}
\begin{barticle}
\bauthor{\bsnm{{Wang}}, \binits{T.}},
\bauthor{\bsnm{{Ofman}}, \binits{L.}},
\bauthor{\bsnm{{Davila}}, \binits{J.M.}}:
\byear{2013},
\batitle{{Three-dimensional Magnetohydrodynamic Modeling of Propagating
  Disturbances in Fan-like Coronal Loops}}.
\bjtitle{\apjl}
\bvolume{775}(\bissue{1}),
\bfpage{L23}.
\doiurl{https://doi.org/10.1088/2041-8205/775/1/L23}.
\adsurl{2013ApJ...775L..23W}.
\end{barticle}
\endbibitem

\bibitem[\protect\citeauthoryear{{Wang}}{2016}]{2016GMS...216..395W}
\begin{barticle}
\bauthor{\bsnm{{Wang}}, \binits{T.J.}}:
\byear{2016},
\batitle{{Waves in Solar Coronal Loops}}.
\bjtitle{Washington DC American Geophysical Union Geophysical Monograph Series}
\bvolume{216},
\bfpage{395}.
\doiurl{https://doi.org/10.1002/9781119055006.ch23}.
\adsurl{2016GMS...216..395W}.
\end{barticle}
\endbibitem

\bibitem[\protect\citeauthoryear{{Wang}
  \textit{et~al.}}{2015}]{2015ApJ...811L..13W}
\begin{barticle}
\bauthor{\bsnm{{Wang}}, \binits{T.}},
\bauthor{\bsnm{{Ofman}}, \binits{L.}},
\bauthor{\bsnm{{Sun}}, \binits{X.}},
\bauthor{\bsnm{{Provornikova}}, \binits{E.}},
\bauthor{\bsnm{{Davila}}, \binits{J.M.}}:
\byear{2015},
\batitle{{Evidence of Thermal Conduction Suppression in a Solar Flaring Loop by
  Coronal Seismology of Slow-mode Waves}}.
\bjtitle{\apjl}
\bvolume{811}(\bissue{1}),
\bfpage{L13}.
\doiurl{https://doi.org/10.1088/2041-8205/811/1/L13}.
\adsurl{2015ApJ...811L..13W}.
\end{barticle}
\endbibitem

\bibitem[\protect\citeauthoryear{{Wang}
  \textit{et~al.}}{2018}]{2018ApJ...860..107W}
\begin{barticle}
\bauthor{\bsnm{{Wang}}, \binits{T.}},
\bauthor{\bsnm{{Ofman}}, \binits{L.}},
\bauthor{\bsnm{{Sun}}, \binits{X.}},
\bauthor{\bsnm{{Solanki}}, \binits{S.K.}},
\bauthor{\bsnm{{Davila}}, \binits{J.M.}}:
\byear{2018},
\batitle{{Effect of Transport Coefficients on Excitation of Flare-induced
  Standing Slow-mode Waves in Coronal Loops}}.
\bjtitle{\apj}
\bvolume{860}(\bissue{2}),
\bfpage{107}.
\doiurl{https://doi.org/10.3847/1538-4357/aac38a}.
\adsurl{2018ApJ...860..107W}.
\end{barticle}
\endbibitem

\bibitem[\protect\citeauthoryear{{Wang}
  \textit{et~al.}}{2021}]{2021SSRv..217...34W}
\begin{barticle}
\bauthor{\bsnm{{Wang}}, \binits{T.}},
\bauthor{\bsnm{{Ofman}}, \binits{L.}},
\bauthor{\bsnm{{Yuan}}, \binits{D.}},
\bauthor{\bsnm{{Reale}}, \binits{F.}},
\bauthor{\bsnm{{Kolotkov}}, \binits{D.Y.}},
\bauthor{\bsnm{{Srivastava}}, \binits{A.K.}}:
\byear{2021},
\batitle{{Slow-Mode Magnetoacoustic Waves in Coronal Loops}}.
\bjtitle{\ssr}
\bvolume{217}(\bissue{2}),
\bfpage{34}.
\doiurl{https://doi.org/10.1007/s11214-021-00811-0}.
\adsurl{2021SSRv..217...34W}.
\end{barticle}
\endbibitem

\bibitem[\protect\citeauthoryear{{Yuan} and
  {Nakariakov}}{2012}]{2012A&A...543A...9Y}
\begin{barticle}
\bauthor{\bsnm{{Yuan}}, \binits{D.}},
\bauthor{\bsnm{{Nakariakov}}, \binits{V.M.}}:
\byear{2012},
\batitle{{Measuring the apparent phase speed of propagating EUV disturbances}}.
\bjtitle{\aap}
\bvolume{543},
\bfpage{A9}.
\doiurl{https://doi.org/10.1051/0004-6361/201218848}.
\adsurl{2012A&A...543A...9Y}.
\end{barticle}
\endbibitem

\bibitem[\protect\citeauthoryear{{Zavershinskii}
  \textit{et~al.}}{2019}]{2019PhPl...26h2113Z}
\begin{barticle}
\bauthor{\bsnm{{Zavershinskii}}, \binits{D.I.}},
\bauthor{\bsnm{{Kolotkov}}, \binits{D.Y.}},
\bauthor{\bsnm{{Nakariakov}}, \binits{V.M.}},
\bauthor{\bsnm{{Molevich}}, \binits{N.E.}},
\bauthor{\bsnm{{Ryashchikov}}, \binits{D.S.}}:
\byear{2019},
\batitle{{Formation of quasi-periodic slow magnetoacoustic wave trains by the
  heating/cooling misbalance}}.
\bjtitle{Physics of Plasmas}
\bvolume{26}(\bissue{8}),
\bfpage{082113}.
\doiurl{https://doi.org/10.1063/1.5115224}.
\adsurl{2019PhPl...26h2113Z}.
\end{barticle}
\endbibitem

\bibitem[\protect\citeauthoryear{{Zavershinskii}
  \textit{et~al.}}{2020}]{2020PhRvE.101d3204Z}
\begin{barticle}
\bauthor{\bsnm{{Zavershinskii}}, \binits{D.I.}},
\bauthor{\bsnm{{Molevich}}, \binits{N.E.}},
\bauthor{\bsnm{{Riashchikov}}, \binits{D.S.}},
\bauthor{\bsnm{{Belov}}, \binits{S.A.}}:
\byear{2020},
\batitle{{Nonlinear magnetoacoustic waves in plasma with isentropic thermal
  instability}}.
\bjtitle{\pre}
\bvolume{101}(\bissue{4}),
\bfpage{043204}.
\doiurl{https://doi.org/10.1103/PhysRevE.101.043204}.
\adsurl{2020PhRvE.101d3204Z}.
\end{barticle}
\endbibitem

\bibitem[\protect\citeauthoryear{{Zavershinsky} and
  {Molevich}}{2013}]{2013TePhL..39..676Z}
\begin{barticle}
\bauthor{\bsnm{{Zavershinsky}}, \binits{D.I.}},
\bauthor{\bsnm{{Molevich}}, \binits{N.E.}}:
\byear{2013},
\batitle{{A magnetoacoustic autowave pulse in a heat-releasing ionized gaseous
  medium}}.
\bjtitle{Technical Physics Letters}
\bvolume{39}(\bissue{8}),
\bfpage{676}.
\doiurl{https://doi.org/10.1134/S1063785013080130}.
\adsurl{2013TePhL..39..676Z}.
\end{barticle}
\endbibitem

\end{thebibliography}

\end{article} 
\end{document}